\documentclass[aps,prd,onecolumn,amsmath,11pt,superscriptaddress,floatfix,nofootinbib,preprintnumbers]{revtex4-2}

\usepackage{mathrsfs}
\usepackage{amsfonts}
\usepackage{amsmath,amssymb,bm}
\usepackage{array}
\usepackage{verbatim}
\usepackage{epsfig}
\usepackage{graphicx}
\usepackage{hyperref}
\hypersetup{colorlinks, linkcolor = [rgb]{0, 0, 0.5}, citecolor = [rgb]{0,0.0,0.5}, urlcolor = [rgb]{0,0.0,0.5}}
\usepackage[normalem]{ulem}
\usepackage{xcolor}
\usepackage{ulem}
\usepackage{multirow}

\usepackage{graphicx}
\graphicspath{{./figs/}}
\usepackage{dcolumn}
\usepackage{bm}
\usepackage{slashed}
\usepackage{siunitx}
\usepackage{ulem,xpatch}
\usepackage{hyperref}
\usepackage[mathlines]{lineno}
\usepackage{comment}
\usepackage{soul}
\usepackage{xcolor}
\usepackage{xfrac}

\allowdisplaybreaks[1]

\begin{document}

\title{Inclusive and semi-inclusive production of spin-3/2 hadrons in $e^+e^-$ annihilation}

\newcommand*{\SDU}{Key Laboratory of Particle Physics and Particle Irradiation (MOE), Institute of Frontier and Interdisciplinary Science, Shandong University, Qingdao, Shandong 266237, China}\affiliation{\SDU}

\author{Jing~Zhao}
\author{Zhe~Zhang}
\author{Zuo-tang~Liang}
\author{Tianbo~Liu}\email{liutb@sdu.edu.cn}
\author{Ya-jin~Zhou}\email{zhouyj@sdu.edu.cn}
%


\begin{abstract}

We investigate the inclusive and semi-inclusive productions of spin-3/2 hadrons, such as $\Omega$, in unpolarized $e^+ e^-$ annihilation. The general differential cross sections are expressed in terms of structure functions in accordance to the polarization of the hadron and the azimuthal modulations. 
We derive a complete definition of quark transverse momentum dependent (TMD) fragmentation functions (FFs) to spin-3/2 hadrons for the first time from the decomposition of the quark-quark correlation matrix at leading twist, 14 of which are newly defined corresponding to rank-3 tensor polarized hadron. The collinear FFs are obtained from the $k_T$-integrated correlation matrix, and only two TMD FFs with rank-3 tensor polarization have nonvanishing collinear counterparts. 
Then we perform a leading order calculation of the unpolarized differential cross sections. In the single-hadron inclusive production, only two structure functions are found nonzero and none of the rank-3 tensor polarized FFs contributes. For the nearly back-to-back two-hadron production, half of the 48 structure functions are found nonzero even if the spin of the second hadron is not analyzed, and ten of the rank-3 tensor polarized TMD FFs contribute. 
Therefore, one can study the rank-3 tensor polarized FFs via the production of a spin-3/2 hadron and an unpolarized hadron in unpolarized $e^+ e^-$ collision experiments. These newly defined FFs can be further applied in semi-inclusive deep inelastic scattering processes for the study of nucleon structures.

\end{abstract}

\maketitle
\section{Introduction}
\label{s.intro}

The quantum chromodynamics (QCD) is nowadays known as the fundamental theory of strong interactions within the framework of the Yang-Mills gauge field theory. The color confinement, as a feature of QCD, prevents the direct measurement of its fundamental degrees of freedom, quarks and gluons, by any modern detectors. To understand the emergence of color neutral hadrons from colored quarks and gluons produced at high-energy scatterings has become one of the main issues in nuclear and particle physics~\cite{Anselmino:1994gn,ALEPH:1996oqp,Aidala:2012mv,Filippone:2001ux,Chen:2015tca,Anselmino:2020vlp,Metz:2016swz,Anderle:2021wcy,ParticleDataGroup:2020ssz}. Due to the nonperturbative nature of QCD at hadronic scales, it is still challenging to calculate the hadronization from first principles, and most analyses of hadron involved high energy reactions rely on the QCD factorization~\cite{Collins:1989gx}. In the factorized formalism, the cross section is approximately expressed as a convolution of the short-distance partonic scattering cross section, which can be calculated order by order in perturbation theory, and some universal process-independent long-distance functions with overall corrections suppressed by inverse powers of some hard scale.  

The fragmentation function (FF), {\it e.g.}, $D_{f\to h}(z)$, was first introduced in QCD to describe inclusive production of a desired hadron in $e^+e^-$ annihilation~\cite{Collins:1981uw}, although the concept arose right after the parton model~\cite{Berman:1971xz}. It describes the probability density of observing the hadron $h$ carrying a fraction $z$ of the longitudinal momentum from a parton with flavor $f$. If an additional hadron is identified in the final state, the reaction will become a double scale process when the transverse momentum imbalance, $|{\bm P}_{1T}+{\bm P}_{2T}|$, of the two hadrons is much smaller than the hard scale $Q=\sqrt{s}$, the center-of-mass energy of the reaction. One needs to apply the transverse momentum dependent (TMD) factorization~\cite{Collins:1981uw,Collins:1981uk,Aybat:2011zv} and the cross section is expressed as the convolution of partonic hard part and two TMD FFs $D_{f\to h}(z, k_{T}^2)$, where $k_T$ is the transverse momentum of the parton with respect to the motion of the observed hadron. Taking the quark spin degree of freedom into account, one can define polarized TMD FFs, which encode the correlation between quark transverse momentum and its spin during the hadronization. They lead to rich phenomena in high energy reactions, such as the azimuthal asymmetries measured by Belle~\cite{Belle:2005dmx,Belle:2008fdv,Belle:2019nve}, BaBar~\cite{BaBar:2013jdt}, and BESIII~\cite{BESIII:2015fyw} in inclusive production of two hadrons in unpolarized $e^+e^-$ annihilation. In recent years, a factorization scheme is also developed to access the TMD FF in the single hadron production process in $e^+ e^-$ annihilation by identifying the thrust axis, which provides the sensitivity to the transverse momentum of the detected hadron~\cite{Kang:2020yqw,Boglione:2020auc,Boglione:2020cwn,Makris:2020ltr}. Corresponding phenomenological studies based on the data from Belle~\cite{Belle:2018ttu,Belle:2019ywy} have been carried out~\cite{Boglione:2022nzq,Gamberg:2021iat,Boglione:2021wov,Soleymaninia:2019jqo}. While this process provides us new opportunities to learn TMD FFs, it is beyond the scope of this study. In the rest of this paper, we only focus on inclusive and semi-inclusive processes as described above without identifying the thrust axis.

The TMD FFs extracted from $e^+e^-$ annihilation also play an important role in the study of nucleon structures through the semi-inclusive deep inelastic scattering (SIDIS) process. For instance, the Collins fragmentation function $H_{1}^{\perp}(z,k_T^2)$~\cite{Collins:1992kk}, which is naively interpreted as the probability density of transversely polarized quark fragmenting to unpolarized hadron, can lead to a target single spin asymmetry by convoluting with quark transversity distribution, which describes the net density of a transversely polarized quark in a transversely polarized nucleon. This azimuthal asymmetry, usually referred to as the Collins asymmetry~\cite{HERMES:2004mhh}, is currently the main observable for phenomenological extraction of quark transversity distributions of the nucleon. Besides, by selecting different type of hadrons, such as pions and kaons, in the SIDIS process, one may have flavor separation in the extraction of the partonic structures of the nucleon~\cite{Sato:2019yez}.

The spin, as a fundamental property of all particles, is proven a powerful quantity to test theories and models, especially in hadron physics. Although the spin effect in the hadronization process exists even in the fragmentation to spin-0 or unpolarized hadrons, such as the Collins effect~\cite{Collins:1992kk, Boer:2003cm}, much richer information can be obtained from the measurement of the spin state of the hadron~\cite{Mulders:1995dh,Boer:1999uu,Goeke:2005hb,Boer:1997mf,Boer:2008fr,Vogelsang:2005cs,Pitonyak:2013dsu,Garzia:2016kqk,Yang:2016qsf,Chen:2021zrr}. The recent measurement of the spontaneous polarization of $\Lambda$ from unpolarized $e^+e^-$ collisions by Belle has received great interest~\cite{Belle:2018ttu}. This observation is suggested to be explained by a naively time-reversal-odd (T-odd) FF $D_{1T}^\perp(z,k_T^2)$ and several parametrizations have been proposed by fitting the Belle data~\cite{DAlesio:2020wjq,Callos:2020qtu,Chen:2021hdn}. The polarized FFs to $\Lambda$ also provide opportunities to study nucleon spin structures via the spin transfer in SIDIS process~\cite{Ellis:1995fc,Lu:1995np,Chi:2013hka,Chi:2014xba,Du:2016irt,Du:2017nzy} with particular sensitivity to the spin of strange quarks. Extending to vector mesons, such as $\rho$ and $K^*$, there are five additional spin states, referred to as rank-2 tensor polarization~\cite{Bourrely:1980mr,Leader:2011vwq,Bacchetta:2000jk,Biedenharn:1958}, from the decomposition of the spin density matrix. The measurements of the rank-2 tensor polarization of vector mesons produced in $e^+e^-$~\cite{DELPHI:1997ruo,OPAL:1997vmw,OPAL:1999hxs,OPAL:1999hxs,OPAL:1997nwj} and $pp$~\cite{NOMAD:2006kuc,Chen:2008zzg,STAR:2008lcm,Zhou:2019lun,ALICE:2019aid,Chen:2008zzg,Casula:2013mla,Singh:2018uad,Kundu:2021lra,STAR:2022fan} collisions have been carried out and the results incite extensive theoretical studies~\cite{Bacchetta:2000jk,Mulders:2000sh,Bacchetta:2001rb,Gliske:2014wba,Boer:2016xqr,Ninomiya:2017ggn,Kumano:2020ijt,Cotogno:2017puy,Kumano:2019igu,Wei:2013csa,Wei:2014pma,Chen:2016moq,Chen:2020pty,Jiao:2022gzu,Kumano:2021xau,Kumano:2021fem}. A systematic study of the Lorentz decomposition of the quark-quark correlator suggests that the spin alignment of vector mesons is solely determined by a rank-2 tensor polarized FF $D_{1LL}(z)$ independent of the spin of the fragmenting quark~\cite{Chen:2016iey} and some parametrizations have been obtained by fitting the LEP data~\cite{DELPHI:1997ruo,OPAL:1997vmw,OPAL:1999hxs}. 

In this paper, we present a systematic study of the TMD FFs for spin-3/2 baryons at leading twist. Apart from the three components of the vector polarization and the five components of the rank-2 tensor polarization, we define the seven components of the rank-3 tensor polarization from the decomposition of the spin density matrix, while keeping the orthogonal relation among different spin projectors. Then we derive the complete set of TMD FFs at leading twist via Lorentz decompositions of the quark-quark correlator and 14 of them are related to the rank-3 tensor polarization. The corresponding collinear FFs are obtained from the $k_T$-integrated quark-quark correlator and only two of the rank-3 tensor polarized FFs have collinear counter parts. As an ideal process to investigate FFs, we study the production of spin-3/2 hadrons in unpolarized $e^+e-$ annihilation. For the inclusive production, we derive the differential cross section in terms of nine structure functions in accordance to different angular dependence and polarization of the hadron. As a single scale process, we apply the collinear factorization formalism. Expressing the structure functions in terms of FFs, we find that none of the rank-3 tensor polarized FFs contributes at leading order. 

To access the TMD FFs, it is necessary to have a second scale, which can be achieved by measuring one additional hadron. Since the measurement of the spin states of the two hadrons is very challenging, we consider the inclusive production of a spin-3/2 hadron and an unpolarized hadron in unpolarized $e^+e^-$ annihilation. The differential cross section can be generally expressed in terms of 48 structure functions. Within the kinematic region where the two hadrons are nearly back to back, we apply the TMD factorization formalism and perform a leading order calculation. At leading twist, 24 of the structure functions are found nonzero as convolutions of two TMD FFs and ten of the structure functions have contributions from the rank-3 tensor polarized TMD FFs. Similar to the vector polarization of the baryon, the rank-3 tensor polarization of spin-3/2 hadrons can, in principle, be measured via weak decays, and hence these newly defined rank-3 tensor polarized TMD FFs can be explored via $e^+e^-\to \Omega h X$ at Belle, BaBar, and BESIII experiments. 

The paper is organized as follows. In Sec.~\ref{s.density matrix}, we provide a decomposition of the spin density matrix of spin-3/2 hadrons. In Sec.~\ref{s.TMD Fragmentation functions}, we derive the TMD FFs and their collinear counter parts at leading twist from the quark-quark correlation function. The general kinematic analysis of the differential cross section and the calculation of the structure functions in terms of the FFs in the parton model are presented in Sec.~\ref{s.inclusive} for the inclusive production process and in Sec.~\ref{s.semi-inclusive} for the semi-inclusive production process. A summary and outlook is given in Sec.~\ref{s.summary}.

\section{Spin density matrix of spin-3/2 particles} \label{s.density matrix}

The complete information of a quantum system is given by the density matrix, which is capable of describing both pure states and mixed states. Focusing on the spin degree of freedom, one can use the spin density matrix to characterize the spin state of a system. Since a massive spin-$s$ particle belongs to the $(2s + 1)$-dimensional representation of the $SO(3)$ group, the corresponding spin density matrix $\rho$ is a $(2s + 1) \times (2s + 1)$ matrix. 

We start from the spin density matrix for spin-3/2 particles, which has been derived in the study of nucleon resonance~\cite{Song:1967}. According to the Hermiticity condition $\rho = \rho^\dagger$, one can decompose $\rho$ on a basis of 16 independent Hermitian matrices,  
\begin{align}
\rho=\frac{1}{4}\left(\bm{1}+\frac{4}{5} S^{i} \Sigma^{i}+\frac{2}{3} T^{i j} \Sigma^{i j}+\frac{8}{9}R^{i j k}\Sigma^{i j k}\right).  \label{e.density}  
\end{align}
Here the $(\Sigma^a)$s with $a=i$, $ij$, and $ijk$ are traceless matrices, corresponding to vector polarization, rank-2 tensor polarization, and rank-3 tensor polarization states, respectively. When summing over all spin states, only the scalar term $\bm 1$ contributes and the normalization condition ${\rm Tr[\rho]=1}$ is satisfied. 
On the $s_z$ basis, the three $(\Sigma^i)$s are represented as
\begin{equation}
\Sigma^x=
    \begin{pmatrix}
    0 & \frac{\sqrt{3}}{2} & 0 & 0 \\
    \frac{\sqrt{3}}{2} & 0 & 1 & 0 \\
    0 & 1 & 0 & \frac{\sqrt{3}}{2} \\
    0 & 0 & \frac{\sqrt{3}}{2} & 0 \\
    \end{pmatrix},
\quad
\Sigma^y=
    \begin{pmatrix}
    0 & -\frac{i\sqrt{3}}{2} & 0 & 0 \\
    \frac{i\sqrt{3}}{2} & 0 & -i & 0 \\
    0 & i & 0 & -\frac{i\sqrt{3}}{2} \\
    0 & 0 & \frac{i\sqrt{3}}{2} & 0 \\
    \end{pmatrix},
\quad
\Sigma^z=
    \begin{pmatrix}
    \frac{3}{2} & 0 & 0 & 0 \\
    0 & \frac{1}{2} & 0 & 0 \\
    0 & 0 & -\frac{1}{2} & 0 \\
    0 & 0 & 0 & -\frac{3}{2} \\ 
    \end{pmatrix}.
\end{equation}
They form a representation of the angular momentum algebra, $[\Sigma^i, \Sigma^j]=i\epsilon^{ijk}\Sigma^k$, and the Casimir operator is given by $\Sigma^x \Sigma^x + \Sigma^y \Sigma^y + \Sigma^z \Sigma^z = \frac{15}{4} {\bm 1}$.
The five rank-2 tensor polarization basis $(\Sigma^{ij})$s can be constructed from $(\Sigma^i)$s,
\begin{align}
\Sigma^{i j}&=\frac{1}{2}\left(\Sigma^{i} \Sigma^{j}+\Sigma^{j} \Sigma^{i}\right)-\frac{5}{4} \delta^{i j}{\bm 1},
\label{e.Sigmaij}
\end{align}
where the second term subtracts the scalar component from the product in the first term and one has the relation
\begin{align}
    \Sigma^{x x}+\Sigma^{y y}+\Sigma^{z z}&=0.
\end{align}
Similarly, we construct the seven rank-3 tensor polarization basis $(\Sigma^{ijk})$s as
\begin{align}
    \Sigma^{i j k}&=\frac{1}{6} \Sigma^{\{i} \Sigma^{j}\Sigma^{k\}}
    -\frac{41}{60}  \left(\delta^{ij}\Sigma^k + \delta^{jk}\Sigma^i + \delta^{ki}\Sigma^j\right) \notag \\
    &= \frac{1}{3}\left(\Sigma^{ij}\Sigma^{k} + \Sigma^{jk}\Sigma^{i} + \Sigma^{ki}\Sigma^{j} \right) 
    - \frac{4}{15}\left(\delta^{ij}\Sigma^k + \delta^{jk}\Sigma^i + \delta^{ki}\Sigma^j \right),
    \label{e.Sigmaijk}
\end{align}
where the symbol $\{\cdots\}$ stands for a symmetrization of the indices, {\it i.e.,} the sum over all permutations. The second term substracts the vector components from the product in the first term and one can derive the relations
\begin{align} 
 \Sigma^{x x z}+\Sigma^{y y z}+\Sigma^{z z z}&=0,\\
 \Sigma^{x x y}+\Sigma^{y y y}+\Sigma^{z z y}&=0,\\
 \Sigma^{x x x}+\Sigma^{y y x}+\Sigma^{z z x}&=0.
\end{align}
The basis matrices constructed via the above procedure automatically satisfy the orthogonal relation
\begin{align}
    {\rm Tr}[\Sigma^i \Sigma^{jk}] = {\rm Tr}[\Sigma^i \Sigma^{jkl}] = {\rm Tr}[\Sigma^{ij} \Sigma^{klm}] = 0,
\end{align}
while one still has the freedom for the choice of basis matrices within each subspace. For convenience, we impose additional orthogonal conditions for the choice of the basis, requiring ${\rm Tr}[\Sigma^a \Sigma^b] = 0$ if $\Sigma^a \neq \Sigma^b$, where $\Sigma^a={\bm 1}$, $\Sigma^i$, $\Sigma^{ij}$, and $\Sigma^{ijk}$. Then the 16 $(\Sigma^a)$s can serve as spin projectors and the coefficients $S^i$, $T^{ij}$, and $R^{ijk}$ in Eq.~\eqref{e.density} can be calculated from $\langle \Sigma^a \rangle \equiv {\rm Tr}[\rho \Sigma^a]$ with corresponding $\Sigma^a$. 
While the spin density matrices for spin-1/2 and spin-1 are well known and even available in some textbooks~\cite{Bourrely:1980mr,Leader:2011vwq,Bacchetta:2000jk,Biedenharn:1958}, the one for spin-3/2 particles is not widely utilized. Our choice of the decomposition basis matrices, Eqs.~\eqref{e.Sigmaij} and~\eqref{e.Sigmaijk}, are consistent with those in Refs.~\cite{Song:1967,Kim:1976dn,Choi:1989yf} up to a normalization constant, but different from the form in Ref.~\cite{Yang:2020ezt}.
We comment that the basis matrices in Ref.~\cite{Yang:2020ezt} do not satisfy the orthogonal condition and hence should not be directly used as projectors to extract corresponding coefficients.

Following the common convention, we define the spin vector $S^i$ in the hadron rest frame as
\begin{align}
    S^i&=\left(S_{T}^{x}, S_{T}^{y}, S_{L}\right), 
    \label{e.spinvector}
\end{align}
where the components are given by
\begin{align}
    S_{L}=\langle \Sigma^{z} \rangle,
    \quad
    S_{T}^{x}=\langle \Sigma^{x} \rangle,
    \quad
    S_{T}^{y}=\langle \Sigma^{y} \rangle.
    \label{e.components1}
\end{align}
Similarly, we define the rank-2 spin tensor $T^{ij}$ as
\begin{align}
    T^{ij}
    &=\frac{1}{2} 
    \begin{pmatrix}
        -S_{L L}+S_{T T}^{x x} & S_{T T}^{x y} & S_{L T}^{x} \\
        S_{T T}^{x y} &  -S_{L L}-S_{T T}^{x x} & S_{L T}^{y} \\
        S_{L T}^{x} & S_{L T}^{y} & 2S_{L L} 
    \end{pmatrix},
    \label{e.tensor1}
\end{align}
where the components are given by
\begin{align}
 S_{LL}&=\langle \Sigma^{zz} \rangle,
 \quad
 S_{LT}^{x} = 2 \langle \Sigma^{xz} \rangle,
 \quad
 S_{LT}^{y} = 2 \langle \Sigma^{yz} \rangle,
 \nonumber\\
 S_{TT}^{xy}&=2 \langle \Sigma^{xy} \rangle,
 \quad
 S_{TT}^{xx} = \langle \Sigma^{xx}-\Sigma^{yy} \rangle.
 \label{e.components2}
\end{align}
The rank-3 spin tensor $R^{ijk}$ is defined as
\begin{equation}
R^{i j k}=\frac{1}{4}
\left[
\begin{aligned}
&\left(
\begin{array}{ccc}
-3S_{LLT}^x+S_{TTT}^{xxx} & -S_{LLT}^y+S_{TTT}^{yxx} & -2S_{LLL}+S_{LTT}^{xx}\\
 -S_{LLT}^y+S_{TTT}^{yxx} & -S_{LLT}^x-S_{TTT}^{xxx} & S_{LTT}^{xy}\\
 -2S_{LLL}+S_{LTT}^{xx} &S_{LTT}^{xy} & 4S_{LLT}^{x}
\end{array}
\right)\\
&\left(
\begin{array}{ccc}
-S_{LLT}^y+S_{TTT}^{yxx} & -S_{LLT}^x-S_{TTT}^{xxx} & S_{LTT}^{xy}\\
-S_{LLT}^x-S_{TTT}^{xxx} & -3S_{LLT}^y-S_{TTT}^{yxx} & -2S_{LLL}-S_{LTT}^{xx}\\
S_{LTT}^{xy} & -2S_{LLL}-S_{LTT}^{xx} & 4S_{LLT}^{y}
\end{array}
\right)\\
&\left(
\begin{array}{ccc}
 -2S_{LLL}+S_{LTT}^{xx} & S_{LTT}^{xy} & 4S_{LLT}^{x}\\
 S_{LTT}^{xy} & -2S_{LLL}-S_{LTT}^{xx} & 4S_{LLT}^{y}\\
 4S_{LLT}^{x} & 4S_{LLT}^{y} & 4S_{LLL}
\end{array}
\right)
\end{aligned}
\right],
\label{e.tensor2}
\end{equation}
where the components are given by
\begin{align}
 S_{L L L}&=\langle \Sigma^{z z z} \rangle,
 \quad
 S_{L L T}^x =\langle \Sigma^{x z z} \rangle,
 \quad
 S_{L L T}^y =\langle \Sigma^{y z z} \rangle,
 \quad
 S_{L T T}^{xy} =4\langle \Sigma^{x y z} \rangle,
 \nonumber\\
 S_{L T T}^{xx}&=2 \langle \Sigma^{x x z} - \Sigma^{y y z} \rangle,
 \quad
 S_{T T T}^{xxx} = \langle \Sigma^{x x x} - 3 \Sigma^{x y y}\rangle,
 \quad
 S_{T T T}^{yxx} = \langle 3 \Sigma^{y x x} - \Sigma^{y y y}\rangle.
 \label{e.components3}
\end{align}
We should note that one may have different expressions for $S^i$, $T^{ij}$, and $R^{ijk}$ due to the convention of the value range of the components. The value ranges of the spin components defined above and their physical interpretations are provided in Appendix~\ref{a.SDM}.  

One can further define the total degree of polarization from the spin density matrix,
\begin{align}
d=\frac{1}{\sqrt{2 s}} \sqrt{(2 s+1) {\rm Tr}[\rho^{2}]-1}.
\end{align}
According to Eq.~\eqref{e.density}, the total degree of polarization of a spin-3/2 particles is 
\begin{align}
d&=\frac{1}{3}\left\{ \frac{12}{3}S^i S_i +2T^{ij} T_{ij}+\frac{8}{3}R^{ijk}R_{ijk}\right\}^\frac{1}{2}\nonumber\\ 
&=\frac{1}{3}\bigg\{ \frac{12}{5} \Big[(S_L)^2+(S_T^x)^2+(S_T^y)^2 \Big] \nonumber\\ 
&\quad 
+ \Big[3(S_{LL})^2+(S_{LT}^x)^2+(S_{LT}^y)^2+(S_{TT}^{xx})^2+(S_{TT}^{xy})^2\Big] \nonumber\\ 
&\quad +\frac{1}{3}\Big[20(S_{LLL})^2+30\left((S_{LLT}^x)^2+(S_{LLT}^y)^2\right) \nonumber\\ 
&\quad\quad + 3\left((S_{LTT}^{xx})^2+(S_{LTT}^{xy})^2\right)+2\left((S_{TTT}^{xxx})^2+(S_{TTT}^{yxx})^2\right)\Big]\bigg\}^\frac{1}{2}.
\end{align}
Its value ranges between 0 and 1.

The spin vector and tensors for a moving hadron can be obtained via a Lorentz boost from those defined in the rest frame. For convenience, we use the light-cone coordinates of a vector $v^\mu = (v^+, v^-, {\bm v}_\perp)$ with $v^\pm = (v^0 \pm v^3)/\sqrt{2}$ and introduce two null vectors $n^\mu = (0,1,{\bm 0}_\perp)$ and $\bar{n}^\mu = (1,0,{\bm 0}_\perp)$. Following the convention in Ref.~\cite{Boer:1997mf}, we take the large component of the hadron momentum along the ``$-$'' direction and decompose the hadron momentum as $P^\mu = n^\mu (P\cdot \bar{n}) + \bar{n}^\mu\, M^2/(2P\cdot \bar{n})$, where $M$ is the mass of the hadron. 
Then, we can express the spin vector and tensors in Lorentz covariant forms as 
\begin{align}
S^{\mu} &= 
S_{L}\left( \frac{M}{2 P \cdot \bar{n}}\bar{n}^\mu-\frac{P  \cdot \bar{n}}{M} n^\mu\right)
+{S}_{T}^\mu, 
\label{e.spin_s}\\
T^{\mu \nu}&=
\frac{1}{2}\Bigg\{  S_{LL}\Bigg[\frac{1}{2}\left(\frac{M}{P \cdot \bar{n}}\right)^2 \bar{n}^\mu\bar{n}^\nu 
+2\left(\frac{P \cdot \bar{n}}{M}\right)^2 n^\mu n^\nu
-\bar{n}^{\{\mu}n^{\nu\}} + g_{T}^{\mu\nu} \Bigg] \nonumber \\ 
&\quad +\frac{1}{2}\left(\frac{M}{P \cdot \bar{n}}\right)\bar{n}^{\{\mu}S_{LT}^{\nu\}}-\left(\frac{P \cdot \bar{n}}{M}\right)n^{\{\mu}S_{LT}^{\nu\}}+S_{TT}^{\mu\nu} \Bigg\},
\label{e.spin_t}\\
R^{\mu \nu \rho}&=
\frac{1}{4}\Bigg\{S_{LLL}\Bigg[\frac{1}{2}\left(\frac{M}{P \cdot \bar{n}}\right)^3 \bar{n}^\mu\bar{n}^\nu\bar{n}^\rho
-\frac{1}{2}\left(\frac{M}{P \cdot \bar{n}}\right)\left(\bar{n}^{\{\mu} \bar{n}^\nu n^{\rho\}}
-\bar{n}^{\{\mu}g_{T}^{\nu\rho\}} \right)       \nonumber\\
&\quad +\left(\frac{P \cdot \bar{n}}{M}\right)\left(\bar{n}^{\{\mu} n^\nu n^{\rho\}}-n^{\{\mu}g_{T}^{\nu\rho\}} \right)-4\left(\frac{P \cdot \bar{n}}{M}\right)^3 n^{\mu} n^\nu n^{\rho}\Bigg]     \nonumber\\
&\quad +\frac{1}{2}\left(\frac{M}{P \cdot \bar{n}}\right)^2 \bar{n}^{\{\mu} \bar{n}^\nu S_{LLT}^{\rho\}}+2\left(\frac{P \cdot \bar{n}}{M}\right)^2 n^{\{\mu} n^\nu S_{LLT}^{\rho\}}-2\bar{n}^{\{\mu} n^\nu S_{LLT}^{\rho\}}+\frac{1}{2}S_{LLT}^{\{\mu}g_{T}^{\nu\rho\}}        \nonumber\\
&\quad +\frac{1}{4}\left(\frac{M}{P \cdot \bar{n}}\right) \bar{n}^{\{\mu}  S_{LTT}^{\nu\rho\}}-\frac{1}{2}\left(\frac{P \cdot \bar{n}}{M}\right) n^{\{\mu}  S_{LTT}^{\nu\rho\}}+S_{TTT}^{\mu\nu\rho}\Bigg\},
\label{e.spin_r}
\end{align}
where the transverse components of $S_T^\mu$, $S_{LT}^{\mu}$, $S_{TT}^{\mu\nu}$, $S_{LLT}^\mu$, $S_{LTT}^{\mu\nu}$, and $S_{TTT}^{\mu\nu\rho}$ are given by
\begin{align}
    S_T^i &= (S_T^x, S_T^y),
    \quad
     S_{LT}^i = (S_{LT}^x, S_{LT}^y),
    \quad
     S_{LLT}^i = (S_{LLT}^x, S_{LLT}^y),
    \nonumber\\
     S_{TT}^{ij} &=
    \begin{pmatrix}
        S_{TT}^{xx} & S_{TT}^{xy} \\
        S_{TT}^{xy} & -S_{TT}^{xx}
    \end{pmatrix},
    \quad
    S_{LTT}^{ij} = 
    \begin{pmatrix}
        S_{LTT}^{xx} & S_{LTT}^{xy} \\
        S_{LTT}^{xy} & -S_{LTT}^{xx}
    \end{pmatrix},
    \nonumber\\
    S_{TTT}^{ijk} &=
    \left[
    \begin{pmatrix}
        S_{TTT}^{xxx} & S_{TTT}^{yxx}\\
        S_{TTT}^{yxx} & -S_{TTT}^{xxx}
    \end{pmatrix},
    \begin{pmatrix}
        S_{TTT}^{yxx} & -S_{TTT}^{xxx}\\
        -S_{TTT}^{xxx} & -S_{TTT}^{yxx}
    \end{pmatrix}
    \right],
    \label{e.spintransverse}
\end{align}
and their longitudinal components are zeros. 
The transverse metric $g_T^{\mu\nu}$ is defined as
\begin{align}
g_{T}^{\mu \nu} &= g^{\mu \nu}-\bar{n}^{\mu} n^{\nu}-n^{\mu} \bar{n}^{\nu}.
\label{e.gt}
\end{align}
One can also find that the spin tensors $S$, $T$, and $R$ are orthogonal to the momentum $P$ of the hadron,
\begin{align}
    P_{\mu} S^\mu=0,
    \quad 
    P_{\mu}T^{\mu\nu}=0,
    \quad  P_{\mu} R^{\mu\nu\rho}=0,
    \label{e.pdotspin}
\end{align}
because these relations are not altered by the Lorentz boost.

\section{TMD Fragmentation functions}
\label{s.TMD Fragmentation functions}

In this section, we derive quark FFs at leading twist from the decomposition of quark-quark correlation function. 

The matrix element that describes the emergence of a spin-3/2 hadron of momentum $P$ characterized by the spin tensors $S$, $T$, and $R$ from a quark is  
$\langle P, S, T, R, X| \bar{\psi}(0) |0\rangle$,
where $\psi$ is the quark field operator and $X$ stands for undetected hadrons. Then one can define the unintegrated quark-quark correlation function,
\begin{align}
\Delta_{\alpha \beta}\left(k, P, S, T, R\right) =&
\sum_X \int \frac{\mathrm{d}^{4} \xi}{(2 \pi)^{4}} e^{i k \cdot \xi}
\langle 0 | \mathcal{L}(\infty,\xi) \psi_{\alpha}(\xi) | P, S, T, R, X\rangle 
\nonumber\\ 
&\times 
\langle P, S, T, R, X |\bar{\psi}_{\beta}(0) \mathcal{L}^\dagger(\infty,0) | 0 \rangle,
\label{e.correlator}
\end{align}  
where $k$ is the momentum of the quark, $\alpha$ and $\beta$ are Dirac indices, and the symbol $\displaystyle {\sum_X}$ also implies the integration over the momenta of the undetected hadrons labeled by $X$.
The gauge link $\mathcal{L}(y_2,y_1)$, which ensures the gauge invariance of the correlation function, is defined as a path integral
\begin{align}
\mathcal{L}(y_2, y_1)&= {\cal P}\exp\left[-i g \int_{y_1}^{y_2} dy \cdot A(y)\right],
\end{align}
where $g$ is the strong coupling constant and $A(y)$ is the gluon field operator.
The correlation function defined in Eq.~\eqref{e.correlator} fulfills the Hermiticity condition,
\begin{align}
\Delta(k, P, S, T, R)&=\gamma^{0} \Delta^{\dagger}(k, P, S, T, R) \gamma^{0} , 
\label{e.hermiticity}
\end{align}
and the parity invariance,
\begin{align}
\Delta(k, P, S, T, R)&=\gamma^{0} \Delta(\bar{k}, \bar{P},-\bar{S}, \bar{T}, -\bar{R}) \gamma^{0},   \label{e.parity}
\end{align}
where $\bar{k}^\mu$, $\bar{P}^\mu$, $\bar{T}^{\mu\nu}$, and $\bar{R}^{\mu\nu\rho}$ represent a sign flip of all space components. The time-reversal invariance has no additional constraints on the correlation function~\cite{DeRujula:1971nnp,Hagiwara:1982cq,Jaffe:1993xb} because of the final-state interaction or formally the gauge link.

From the explicit Dirac indices in Eq.~\eqref{e.correlator}, one can observe that the correlation function is a $4\times4$ matrix in the Dirac space. It can be expanded on a basis of 16 Dirac $\gamma$ matrices, which we choose as $\bf 1$, $\gamma_5$, $\gamma^\mu$, $\gamma^\mu \gamma_5$, and $i \sigma^{\mu\nu}\gamma_5$. The coefficients are given by the combinations of the momenta $k$, $P$, and the spin tensors $S$, $T$, $R$ from the spin density matrix.  After imposing the Hermiticity~\eqref{e.hermiticity} and the parity invariance~\eqref{e.parity}, we obtain the general decomposition of the quark-quark correlation function~\eqref{e.correlator} for a spin-3/2 hadron as
\begin{align}
\Delta(k, P, S, T, R) &=
 B_{1} M \mathbf{1}
+ B_{2} \slashed{P}
+ B_{3} \slashed{k}
+ \left(\frac{B_{4}}{M} \sigma_{\mu \nu} P^{\mu} k^{\nu}\right)
+ \left(i B_{5} k \cdot S \gamma_{5}\right) \nonumber\\
&\quad 
+ B_{6} M \slashed{S} \gamma_{5}
+ B_{7} \frac{k \cdot S}{M} \slashed{P} \gamma_{5}
+ B_{8} \frac{k \cdot S}{M} \slashed{k} \gamma_{5}
+ i B_{9} \sigma_{\mu \nu} \gamma_{5} S^{\mu} P^{\nu} \nonumber\\
&\quad
+ i B_{10} \sigma_{\mu \nu} \gamma_{5} S^{\mu} k^{\nu}
+ i B_{11} \frac{k \cdot S}{M^{2}} \sigma_{\mu \nu} \gamma_{5} P^{\mu} k^{\nu}
+ \left(B_{12} \frac{\epsilon_{\mu \nu \rho \sigma} \gamma^{\mu} P^{\nu} k^{\rho} S^{\sigma}}{M}\right) \nonumber\\
&\quad 
+ \frac{B_{13}}{M} k_{\mu} k_{\nu} T^{\mu \nu} {\bf 1}
+ \frac{B_{14}}{M^{2}} k_{\mu} k_{\nu} T^{\mu \nu} \slashed{P}
+ \frac{B_{15}}{M^{2}} k_{\mu} k_{\nu} T^{\mu \nu} \slashed k \nonumber\\
&\quad 
+ \left(\frac{B_{16}}{M^{3}} k_{\mu} k_{\nu} T^{\mu \nu} \sigma_{\rho \sigma} P^{\rho} k^{\sigma}\right)
+ B_{17} k_{\mu} T^{\mu \nu} \gamma_{\nu}
+ \left(\frac{B_{18}}{M} \sigma_{\nu \rho} P^{\rho} k_{\mu} T^{\mu \nu}\right) \nonumber\\
&\quad 
+ \left(\frac{B_{19}}{M} \sigma_{\nu \rho} k^{\rho} k_{\mu} T^{\mu \nu}\right)
+ \left(\frac{B_{20}}{M^{2}} \epsilon_{\mu \nu \rho \sigma} \gamma^{\mu} \gamma_{5} P^{\nu} k^{\rho} k_{\tau} T^{\tau \sigma}\right)\nonumber\\
&\quad
+ \left(i\frac{B_{21}}{M^2}k_\mu k_\nu k_\rho R^{\mu \nu \rho}\gamma_5\right)
+ \frac{B_{22}}{M^3}k_\mu k_\nu k_\rho R^{\mu \nu \rho}\slashed{k}\gamma_5
+ \frac{B_{23}}{M^3}k_\mu k_\nu k_\rho R^{\mu \nu \rho}\slashed{P}\gamma_5\nonumber\\
&\quad 
+ i\frac{B_{24}}{M^4} k_\mu k_\nu k_\rho R^{\mu \nu \rho} \sigma_{\tau \lambda}\gamma_5 k^\tau P^\lambda 
+ \frac{B_{25}}{M} k_\mu k_\nu R^{\mu \nu \rho}\gamma_\rho\gamma_5 \nonumber\\
&\quad 
+ i\frac{B_{26}}{M^2}\sigma_{\rho \tau}\gamma_5 k^\tau k_\mu k_\nu R^{\mu \nu \rho} 
+ i\frac{B_{27}}{M^2}\sigma_{\rho \tau} \gamma_5 P^\tau k_\mu k_\nu R^{\mu \nu \rho} \nonumber\\
&\quad 
+ \left(\frac{B_{28}}{M^3}\epsilon_{\mu \nu \rho \sigma}\gamma^\mu  k^\nu P^\rho k_\tau k_\lambda R^{\tau \lambda \sigma} \right) , \label{e.decomposition}
\end{align}
where the Dirac indices are suppressed and $\epsilon_{\mu\nu\rho\sigma}$ is the totally antisymmetric tensor with $\epsilon_{+-12}=1$. The $B_i$'s are Lorentz scalar functions of $k\cdot P$ and $k^2$, and the mass factor $M$ is introduced to balance the dimension. The terms in parentheses are often referred to as naively T-odd functions. The terms in Eq.~\eqref{e.decomposition} can be divided into four categories, including the unpolarized ones which do not contain $S$, $T$, or $R$, the vector polarized ones which contain the spin vector $S$, the rank-2 tensor polarized ones which contain the spin tensor $T$, and the rank-3 tensor polarized ones which contain the spin tensor $R$. The rank-3 tensor polarized terms, $B_{21}$ -- $B_{28}$, are newly defined for spin-3/2 hadrons, while the unpolarized, the vector polarized, and the rank-2 tensor polarized ones also exist in the decomposition of the correlation function of spin-1 hadrons~\cite{Bacchetta:2000jk}.

To deal with quark transverse momentum ${\bm k}_T$ in the fragmentation process, we choose the momentum of the hadron along the longitudinal direction with $P^-$ to be the large component, and $P^+=M^2/(2P^-)$ is fixed by the hadron mass. Correspondingly, we perform a Sudakov decomposition of the quark momentum,
\begin{align}
    k^\mu&=\frac{z\left(k^{2}+\boldsymbol{k}_{T}^{2}\right)}{2 P^{-}}\bar{n}^\mu+\frac{P^{-}}{z}n^\mu + {k}_{T}^\mu,
\end{align}
where $z$ is the longitudinal momentum fraction carried by the hadron from the fragmenting quark and $k_T^\mu$ is the transverse part of the full four-momentum vector $k^\mu$ with $k_T^2 = -{\bm k}_T^2$. By integrating Eq.~\eqref{e.correlator} over $k^+$, or, equivalently $k^2$, we obtain the quark-quark correlation function,
\begin{equation}
 \Delta\left(z, k_{T}\right)=\left.\frac{1}{4 z} \int \mathrm{d} k^{+} \Delta\left(k, P, S, T, R\right)\right|_{k^{-}=\frac{P^{-}}{z}}, \label{e.unintcorrrelator}
\end{equation}
which leads to the definition of quark TMD FFs after the Dirac decomposition as shown in Eq.~\eqref{e.decomposition}. 

Here, we consider $P^-$ as a large momentum component and collect the leading twist, {\it i.e.,} twist-two, TMD FFs,  which can be projected out from the correlator $\Delta(z,k_T)$ by the Dirac matrices $\gamma^-$, $\gamma^- \gamma_5$, and $i\sigma^{i-} \gamma_5$. Then a complete set of quark TMD FFs for spin-3/2 hadrons at leading twist are defined as
\begin{align}
\Delta_{U}\left(z, k_{T}\right)=& \frac{1}{4} \left\{ D_{1}\left(z, k_{T}^{2}\right) \slashed{n} +\left(H_{1}^{\perp}\left(z, k_{T}^{2}\right) \sigma_{\mu \nu} \frac{k_{T}^{\mu}}{M} n^{\nu}\right)\right\},
\label{e.tmdff_u}\\
\Delta_{L}\left(z, k_{T}\right)=& \frac{1}{4} \left\{ G_{1 L}\left(z, k_{T}^{2}\right) S_{L} \gamma_{5} \slashed{n} + H_{1 L}^{\perp}\left(z, k_{T}^{2}\right) S_{L} i \sigma_{\mu \nu} \gamma_{5} n^{\mu} \frac{k_{T}^{\nu}}{M}\right\} , \\
\Delta_{T}\left(z, k_{T}\right)=& \frac{1}{4} \left\{G_{1 T}^\perp\left(z, k_{T}^{2}\right) \frac{\boldsymbol{S}_{T} \cdot \boldsymbol{k}_{T}}{M} \gamma_{5} \slashed{n} + H_{1 T}\left(z, k_{T}^{2}\right) i \sigma_{\mu \nu} \gamma_{5} n^{\mu} S_{T}^{\nu} \right. \nonumber\\
&\left.-H_{1 T}^{\perp}\left(z, k_{T}^{2}\right)  i\sigma_{\mu \nu}\gamma_5 n^\mu \frac{k_T^{\nu\rho}}{M^2} S_{T\rho} +\left(D_{1 T}^{\perp}\left(z, k_{T}^{2}\right) \epsilon_{T}^{ \mu\nu} \frac{k_{T \mu}}{M} S_{T\nu} \slashed{n} \right)\right\},\\
\Delta_{L L}\left(z, k_{T}\right)=& \frac{1}{4} \left\{ D_{1 L L}\left(z, k_{T}^{2}\right) S_{L L} \slashed{n}-\left(H_{1 L L}^{\perp}\left(z, k_{T}^{2}\right) S_{L L} \sigma_{\mu \nu} \frac{k_{T}^{\mu}}{M} n^{\nu} \right)\right\}, \\
\Delta_{L T}\left(z, k_{T}\right)=&\frac{1}{4}\left\{  D_{1 L T}^\perp\left(z, k_{T}^{2}\right) \frac{\boldsymbol{S}_{L T} \cdot \boldsymbol{k}_{T}}{M} \slashed{n}- \left(G_{1 L T}^\perp\left(z, k_{T}^{2}\right) \epsilon_{T}^{\mu \nu}  \frac{k_{T \mu}}{M} S_{L T \nu} \gamma_{5}\slashed{n} \right)\right.\nonumber\\
&\left.-\left(H_{1 L T}\left(z, k_{T}^{2}\right) \sigma_{\mu \nu} S_{L T}^\mu n^{\nu} \right)  +\left(H_{1 L T}^{\perp}\left(z, k_{T}^{2}\right) \sigma_{\mu \nu}  \frac{k_T^{\mu \rho}}{M^2} S_{LT\rho} n^{\nu}\right)\right\} , \\
\Delta_{T T}\left(z, k_{T}\right)=&\frac{1}{4} \left\{ D_{1 T T}^\perp\left(z, k_{T}^{2}\right) S_{TT\mu \nu} \frac{k_T^{\mu\nu}}{M^2} \slashed{n} + \left(G_{1 T T}^\perp\left(z, k_{T}^{2}\right) \epsilon_{T \nu}^{\mu }  \frac{k_{T\mu \rho} }{M_{h}^{2}} S_{T T }^{\nu \rho} \gamma_{5} \slashed{n}\right) \right.\nonumber \\
&\left.- \left(H_{1 T T}^{\perp }\left(z, k_{T}^{2}\right) \sigma_{\mu \nu}  S_{T T}^{\mu\rho} \frac{k_{T\rho}}{M_{h}}  n^{\nu}\right) - \left(H_{1 T T}^{\perp \perp}\left(z, k_{T}^{2}\right)   \sigma_{\mu \nu} \frac{k_T^{\mu\rho\sigma}}{M^3} S_{TT\rho\sigma} n^{\nu} \right)\right\},\\
\Delta_{LLL}(z, k_{T})=&\frac{1}{4} \left\{ G_{1 L L L}\left(z, k_{T}^{2}\right) S_{ L L L} \gamma_5 \slashed{n} + H_{1 L L L}^{\perp}\left(z, k_{T}^{2}\right) S_{ L L L}  i \sigma_{\mu \nu} \gamma_5 n^\mu \frac{ k_{T}^\nu}{M}\right\},\\
\Delta_{LLT}(z, k_{T})=& \frac{1}{4} \left\{ G_{1 L L T}^\perp\left(z, k_{T}^{2}\right) \frac{\boldsymbol{S}_{ L L T} \cdot \boldsymbol{k}_{T}}{M}\gamma_5  \slashed{n} - H_{1LLT}^\perp\left(z, k_{T}^{2}\right)   i\sigma_{\mu \nu}\gamma_5 n^\mu \frac{k_T^{\nu \rho}}{M^2} S_{LLT\rho} \right. \nonumber\\
 &\left.+ H_{1LLT} \left(z, k_{T}^{2}\right) i\sigma_{\mu \nu}\gamma_5 n^\mu S_{LLT}^\nu+\left(D_{1 L L T}^\perp\left(z, k_{T}^{2}\right) \epsilon_{T}^{\mu \nu} \frac{k_{T \mu}}{M} S_{L L T \nu}  \slashed{n} \right)\right\},\\
\Delta_{LTT}(z,k_{T})=& \frac{1}{4} \left\{ G_{1LTT}^\perp\left(z, k_{T}^{2}\right) S_{LTT \mu \nu} \frac{k_T^{\mu \nu}}{M^2}\gamma_5 \slashed{n} + H_{1LTT}^{\perp \perp}\left(z, k_{T}^{2}\right)   i\sigma_{\mu \nu}\gamma_5 n^\mu \frac{k_T^{\nu\rho\sigma}}{M^3} S_{LTT \rho \sigma}\right. \nonumber\\
&\left.+ H_{1LTT}^{\perp} \left(z, k_{T}^{2}\right)  i\sigma_{\mu \nu}\gamma_5 n^\mu S_{LTT}^{\nu \rho} \frac{k_{T \rho}}{M}
 + \left(D_{1 L T T}^\perp\left(z, k_{T}^{2}\right)\epsilon_{T \nu}^{\mu}  \frac{k_{T\mu \rho}}{M^{2}} S_{ L T T}^{ \nu \rho} \slashed{n}  \right)\right\},\\
\Delta_{TTT}(z,k_{T})=&\frac{1}{4} \left\{ G_{1TTT}^\perp\left(z, k_{T}^{2}\right) S_{TTT \mu \nu \rho} \frac{k_T^{\mu \nu \rho}}{M^3}\gamma_5 \slashed{n} +H_{1TTT}^{\perp\perp}\left(z, k_{T}^{2}\right)  i\sigma_{\mu \nu}\gamma_5 n^\mu \frac{k_T^{\nu\rho\sigma\tau}}{M^4} S_{TTT \rho \sigma \tau}\right. \nonumber\\
&\left.+H_{1TTT}^\perp \left(z, k_{T}^{2}\right)  i\sigma_{\mu \nu}\gamma_5 n^\mu S_{TTT}^{\nu \rho \sigma} \frac{k_{T \rho \sigma}}{M^2} +\left(D_{1 T T T}^\perp\left(z, k_{T}^{2}\right)\epsilon_{T \nu}^{\mu}  \frac{k_{T\mu \rho \sigma}}{M^{3}} S_{ T T T}^{ \nu \rho \sigma} \slashed{n}\right)\right\},
\label{e.tmdff_ttt}
\end{align}
where $\boldsymbol{S}_{T} \cdot \boldsymbol{k}_{T}=-S_T^\mu k_{T\mu}$, $\boldsymbol{S}_{LT} \cdot \boldsymbol{k}_{T}=-S_{LT}^\mu k_{T\mu}$, and $\boldsymbol{S}_{LLT} \cdot \boldsymbol{k}_{T}=-S_{LLT}^\mu k_{T\mu}$. The terms corresponding to different spin components of the hadron have been separated and labeled by $U$ for unpolarized hadron, by $L$ and $T$ for vector polarized hadron as defined in Eq.~\eqref{e.components1}, by $LL$, $LT$, and $TT$ for rank-2 tensor polarized hadron as defined in Eq.~\eqref{e.components2}, and by $LLL$, $LLT$, $LTT$, and $TTT$ for rank-3 tensor polarized hadron as defined in Eq.~\eqref{e.components3}. The transverse spin tensors $S_T$, $S_{LT}$, $S_{TT}$, $S_{LLT}$, $S_{LTT}$, and $S_{TTT}$ are explicitly written in Eq.~\eqref{e.spintransverse}. The transverse antisymmetric tensor is defined as $\epsilon_T^{\mu\nu}=\epsilon^{\mu\nu\rho\sigma} \bar{n}_{\rho} n_{\sigma}$.
We name the TMD FFs following the common convention. The $D$, $G$, and $H$ are used to represent unpolarized, longitudinally polarized, and transversely polarized quarks, corresponding to the Dirac matrices $\gamma^-$, $\gamma^- \gamma_5$, and $i\sigma^{i-}\gamma_5$, respectively. The subscript ``1'' indicates that the TMD FFs are at leading twist. A superscript ``$\perp$'' is assigned if there is inhomogeneous dependence on the quark transverse momentum, {\it i.e.,} $k_T^\mu$, $k_T^{\mu\nu}$, $k_T^{\mu\nu\rho}$, and $k_T^{\mu\nu\rho\sigma}$, which are completely symmetric and traceless tensors~\cite{Boer:2016xqr} as defined in~Eqs.~\eqref{e.kij}--\eqref{e.kijkl}. The superscript ``$\perp\perp$'' in $H_{1TT}^{\perp\perp}$, $H_{1LTT}^{\perp\perp}$, and $H_{1TTT}^{\perp\perp}$ is introduced to differentiate them from $H_{1TT}^{\perp}$, $H_{1LTT}^{\perp}$, and $H_{1TTT}^{\perp}$.

Among the 32 TMD FFs at leading twist, two are for the unpolarized hadron state, six are for the vector polarized hadron state, ten are for the rank-2 tensor polarized hadron state, and 14 are for the rank-3 tensor polarized hadron state. While the unpolarized, the vector polarized, and the rank-2 tensor polarized ones have been defined in the study of spin-1 hadrons, the rank-3 tensor polarized TMD FFs are newly defined and only exist for hadrons with $s \ge 3/2$. The procedure presented above can be extended to the analysis of hadrons with higher spins. For a rank-$n$ tensor polarized hadron state, one can define $4n+2$ TMD FFs at leading twist.

The collinear FFs are defined from the $k_T$-integrated correlation function,
\begin{equation}
    \Delta(z)=\left.\frac{z}{4} \int \mathrm{d}^{2} \boldsymbol{k}_{T} \mathrm{~d} k^{+} \Delta\left(k, P, S, T, R\right)\right|_{k^{-}=\frac{P_{h}^{-}}{z}}.
    \label{e.deltaz}
\end{equation}
The terms with inhomogeneous $k_T$ dependence in Eqs.~\eqref{e.tmdff_u}--\eqref{e.tmdff_ttt} all vanish and a complete set of collinear FFs are defined as
\begin{align}
\Delta_{U}(z )&=\frac{1}{4} D_1\left(z \right) \slashed{n}, \label{e.deltaU}\\
\Delta_{L}(z )&=\frac{1}{4} G_{1L} \left(z \right) S_L \gamma_5 \slashed{n},\\
\Delta_{T}(z )&=\frac{1}{4} H_{1 T}\left(z \right) i \sigma_{\mu \nu} \gamma_{5} n^{\mu} S_{T}^{\nu},\\
\Delta_{LL}(z )&=\frac{1}{4} D_{1LL}\left(z \right) S_{LL} \slashed{n},\\
\Delta_{LT}(z )&=\frac{1}{4} H_{1 LT}\left(z \right)  \sigma_{\mu \nu} S_{LT}^{\mu} n^{\nu} ,\\
\Delta_{LLL}(z )&=\frac{1}{4} G_{1LLL}\left(z \right)S_{LLL} \gamma
_5\slashed{n}\label{e.deltaLLL},\\
\Delta_{LLT}(z )&=\frac{1}{4} H_{1 LLT}\left(z \right) i \sigma_{\mu \nu} \gamma_{5} n^{\mu} S_{LLT}^{\nu} \label{e.deltaLLT},
\end{align}
where we keep the same symbols as their TMD counterparts. Among the seven collinear FFs, $G_{1LLL}(z)$ and $H_{1LLT}(z)$ are newly defined for the rank-3 tensor polarized hadron state and only exist for hadrons with $s \ge 3/2$, while the other ones have the same definitions as those for a spin-1 hadron~\cite{Bacchetta:2000jk}.

We should note that recovering the collinear FFs from TMD FFs is actually a nontrivial issue in the factorization framework. In the Collins-Soper-Sterman (CSS) formalism~\cite{Collins:1981uk,Collins:1981uw,Collins:1984kg,Collins:2011zzd}, the differential cross section is expressed into a W term and a Y term, also referred to as the ``W+Y'' prescription. The W term dominates the small transverse momentum region where the TMD factorization works. The Y term contains the fixed-order term which dominates the large transverse momentum region within the collinear factorization and the asymptotic term that matches the regions. Since the TMD factorization is not suitable for large transverse momentum, the direct integration of the W term over the transverse momentum to infinity will vanish, and thus the collinear FFs are not directly obtained by integrating the TMD FFs. Various methods have been proposed in literature to deal with this problem and related issues~\cite{Collins:2011zzd,Arnold:1990yk,Bozzi:2005wk,Parisi:1979se,Boer:2015uqa,Boer:2014tka,Collins:2016hqq,Nadolsky:1999kb,Balazs:1997xd}. To suppress the W term at large transverse momentum, one may directly impose a cutoff at some transverse momentum value~\cite{Collins:2011zzd,Collins:2016hqq}. Alternatively, one can work in the Fourier conjugate space, the $b$ space where the resummation is usually implemented, and introduce a cutoff at some small $b$ value~\cite{Arnold:1990yk,Bozzi:2005wk,Parisi:1979se,Boer:2015uqa,Boer:2014tka}. Besides, it is also useful to make the W term short circuited above some large transverse momentum for the computing efficiency~\cite{Nadolsky:1999kb,Balazs:1997xd}. On the other hand, an effective cutoff of the Y term in the small transverse momentum region need to be introduced accordingly~\cite{Collins:1984kg,deFlorian:2001zd}. Therefore, the integration over transverse momentum is a nontrivial issue for a consistent factorization framework and appropriate regularization is required.

\section{Production of spin-3/2 hadrons in $e^{+}e^{-}$ annihilation}
\label{sec:process}

The cleanest process to study FFs is the inclusive hadron production from $e^+ e^-$ annihilation at a large center-of-mass energy $\sqrt{s} = Q$, which serves as a hard probe allowing us to use the parton language. Within the QCD factorization, the cross section can be approximately expressed as a convolution of the production of quarks at short distance and quark FFs to the desired hadron in a certain spin state and overall corrections are suppressed by inverse powers of the hard scale. The inclusive production of a single spin-3/2 hadron, $e^+e^- \to h X$, is characterized by a single hard scale $Q$ and thus not sensitive to the confined motion of quarks during the fragmentation. To have the sensitivity to quark transverse momentum, we can consider the inclusive production of two nearly back to back hadrons, $e^+e^-\to h_1 h_2 X$, where the transverse momentum imbalance of the two hadrons provides a soft scale in addition to the hard scale. As a double-scale process, one should apply the TMD factorization in which the cross section is expressed as the convolution of the production of quarks at short distance and two quark TMD FFs to the detected hadrons $h_1$ and $h_2$.

The spin states of the produced hadron can be measured by analyzing the spatial distribution of the decayed particles. While the strong decays can only be utilized to measure the rank-2 tensor polarization because of the parity conservation, one may extract the vector and rank-3 tensor polarization through weak decays, which have been widely applied in the analysis of $\Lambda$ hyperon polarization. Therefore, to access the rank-3 tensor polarization of the produced hadron, one can choose the identified hadron as the $\Omega$, which only has weak decays.

In this section, we consider both inclusive and semi-inclusive productions of the spin-$3/2$ hadron in $e^+e^-$ annihilation processes. Considering the existing electron-positron colliders, PEP-II, KEKB, and BEPC, the calculation is restricted to unpolarized lepton beams. We first derive the general cross sections in terms of structure functions, which correspond to different angular distributions and spin states of the hadron. At leading twist, we then express the structure functions in terms of the FFs as defined in Sec.~\ref{s.TMD Fragmentation functions}. 

\subsection{Inclusive production of the $\Omega$ in unpolarized $e^+ e^-$ collisions} \label{s.inclusive}

\begin{figure}[ht]
    \centering %
    \includegraphics[width=0.35\textwidth]{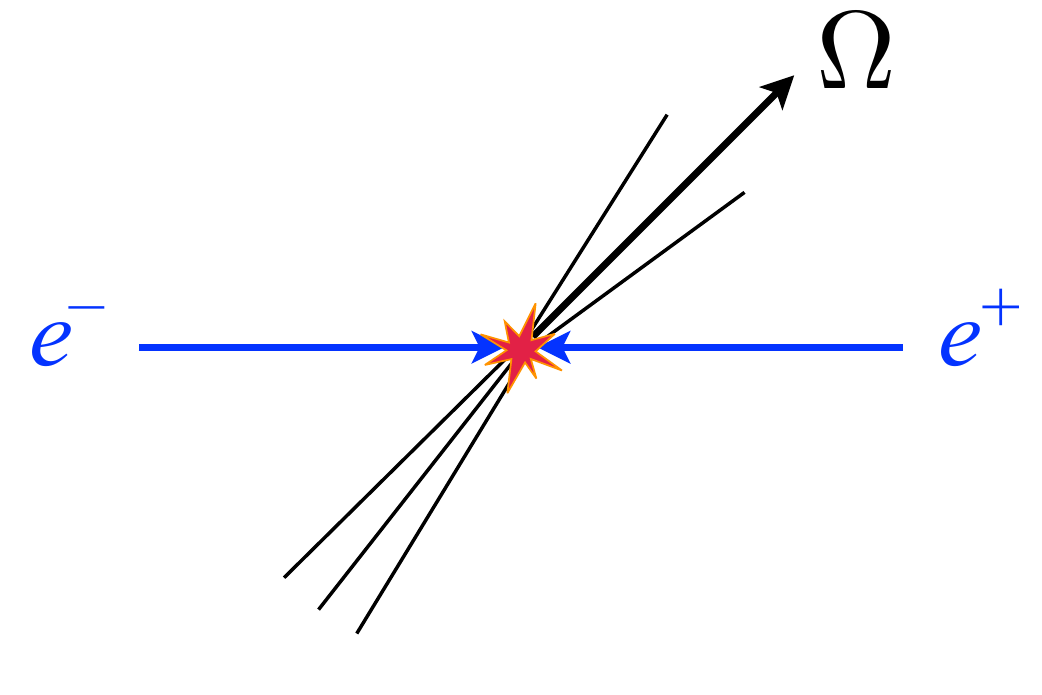}
    \hspace{1cm}
    \includegraphics[width=0.32\textwidth]{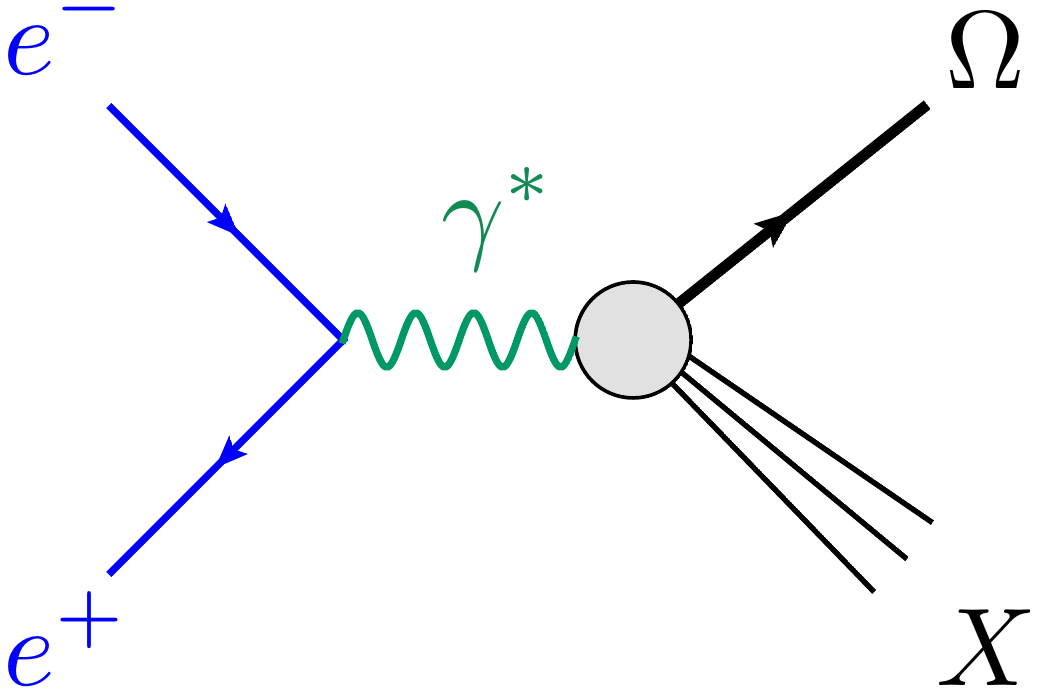}
    \caption{{\bf Left:} The illustration of the inclusive production of $\Omega$ in $e^{+} e^{-}$ annihilation. {\bf Right:} The diagram of the process $e^+e^- \to \Omega X$ under the one-photon-exchange approximation.}
\label{f.omegaX}
\end{figure}

We consider the process
\begin{align}
    e^- (l_1) + e^+ (l_2) \to \Omega (P) + X (P_X),
\end{align}
where the variables in parentheses label the four momenta of the corresponding particles. With one-photon-exchange approximation as illustrated in Fig.~\ref{f.omegaX}, one can write the differential cross section as
\begin{align}
P^{0} \frac{d \sigma}{d^{3} \bm{P}}=\frac{\alpha^{2}}{2 Q^{6}} L_{\mu \nu} W^{\mu \nu}, 
\label{e.dsigma1}
\end{align}
where $\alpha$ is the electromagnetic fine structure constant,
\begin{align}
L_{\mu \nu}\left(l_1, l_2 \right)&=2 [l_{1\mu} l_{2\nu} + l_{1\nu} l_{2\mu} -  g_{\mu \nu}(l_1 \cdot l_2)]
\label{e.lepton}
\end{align}
is the leptonic tensor, and
\begin{align}
W^{\mu \nu}\left(q ; P,S,T,R\right)=&\frac{1}{(2 \pi)} \sum_X \left\langle 0\left|J^{\mu}(0)\right| P_{X} ; P , S,T, R\right\rangle\left\langle P_{X} ; P,  S, T, R\left|J^{\nu}(0)\right| 0\right\rangle \nonumber\\
&\times (2 \pi)^{4} \delta^{(4)}\left(q-P_{X}-P \right)\label{e.hadrontensor}
\end{align}
is the hadronic tensor. Here $q = l_1 + l_2$ is the momentum of the virtual photon with $q^2 = Q^2$, $J^\mu$ is the electromagnetic current operator, and the symbol $\displaystyle{\sum_X}$ represents the sum over the hadronic states $X$ implying the momentum integrals. 

For the unpolarized process, we only need the symmetric part of the hadronic tensor $W^{\mu\nu}$, which satisfies the Hermiticity and the parity invariance relations,
\begin{align}
&W^{* \mu \nu}\left( q ; P,S,T,R\right)=W^{\nu \mu}\left( q ; P,S,T,R\right), \\ 
&W^{\mu \nu}\left( q ; P,S,T,R\right)=W_{\mu \nu}\left(q ; \bar{P},-\bar{S},\bar{T},-\bar{R}\right). 
\end{align}
Besides, the gauge invariance requires $q_\mu W^{\mu\nu} = W^{\mu\nu} q_\nu = 0$. To impose this condition, it is convenient to introduce the so-called conserved vectors and tensors,
\begin{align}
\widetilde{g}^{\mu\nu} &= g^{\mu\nu} - \frac{q^\mu q^\nu}{q^2}, \label{e.gtilde}\\
\widetilde{P}^\mu &= P^\mu - \frac{P\cdot q}{q^2} q^\mu ,\label{e.Ptilde}\\ 
\widetilde{T}^{\mu q}&=T^{\mu q}-\frac{T^{qq}}{q^2}q^\mu, \label{e.tmuq}\\
\widetilde{T}^{\mu \nu}&=T^{\mu \nu}-\frac{T^{\mu q}q^\nu}{q^2}-\frac{T^{\nu q}q^\mu}{q^2}+\frac{T^{qq}q^\mu q^\nu}{q^4},\\
\widetilde{R}^{\mu\nu q} &= R^{\mu\nu q} - \frac{R^{\mu q q} q^\nu}{q^2} - \frac{R^{\nu q q} q^\mu}{q^2} + \frac{R^{qqq} q^\mu q^\nu}{q^4},\label{e.rtilde}
\end{align}
which vanish when contract with $q$. Here, we have used the short-handed notation $A^{\mu q} \equiv A^{\mu \nu} q_\nu$ for conciseness. Then, we can expand the hadronic tensor $W^{\mu\nu}$ as
\begin{align}
W^{  \mu \nu}=\sum_{i=1}^{2} V_{U, i}  t_{U, i}^{  \mu \nu} +
\sum_{i=1}^{1} V_{V, i}  t_{V, i}^{  \mu \nu} + \sum_{i=1}^{4} V_{T, i}  t_{T, i}^{  \mu \nu}+\sum_{i=1}^{2} V_{R, i}  t_{R, i}^{  \mu \nu}, \label{e.wmunu}
\end{align}
where the nine basis tensors are
\begin{align}
t_{U}^{  \mu \nu}&= \left\{\widetilde{g}^{\mu \nu}, \widetilde{P}^{\mu} \widetilde{P}^{\nu} \right\},\\
t_{V}^{  \mu \nu}&= \epsilon^{S P q\{\mu}\widetilde{P}^{\nu\}},\\
t_{T}^{  \mu \nu}&=\left\{T^{qq} \widetilde{g}^{\mu \nu}, T^{qq} \widetilde{P}^{\mu} \widetilde{P}^{\nu}, \widetilde{T}^{q\{\mu}\widetilde{P}^{\nu\}}, \widetilde{T}^{\mu \nu}  \right\},\\
t_{R}^{  \mu \nu}&=\left\{\epsilon^{R^{qq} P q\{\mu} \widetilde{P}^{\nu\}},  R_{\rho}^{~q\{\mu}\epsilon^{\nu\} \rho P q}  \right\},
\end{align}
with the subscripts $U$, $V$, $T$, and $R$ indicating the polarization of the hadron $\Omega$. The coefficients $V_{U,i}$, $V_{V,i}$, $V_{T,i}$, and $V_{R,i}$ are scalar functions of $P\cdot q$ and $q^2$, often referred to as structure functions. Although the final definition of structure functions are linear combinations of these coefficients, the number of independent structure functions is completely determined by the number of basis tensor from this kinematic analysis.

After contracting the leptonic tensor and the hadronic tensor, we can obtain the general differential cross section in Eq.~\eqref{e.dsigma1}. 
Whereas, it is practically convenient to specify a reference frame and to separate the contributions from each structure function according to the angular distribution of the observed hadron as well as its spin states. 
As a natural choice, we consider the differential cross section in the center-of-mass frame of leptons and define the temporal basis vector
\begin{align}
\hat{t}^\mu&=\frac{q^\mu}{Q}.\label{e.that1}
\end{align} 
The spatial vector $\hat{z}$ is chosen opposite to the direction of the hadron momentum and can be written in covariant form as
\begin{align}
\hat{z}^\mu&=\frac{z_h q^\mu-2P^\mu}{z_h Q \sqrt{1-\gamma_h^2}},\label{e.zhat1}
\end{align}
where
\begin{align}
    z_h=\frac{2P \cdot q}{Q^2},
    \quad
    \gamma_h=\frac{2M}{z_h Q}.
\end{align}
We keep the $\gamma_h$ here since the mass of the $\Omega$ may not be negligible.
Following the definitions of the light-cone basis vectors 
\begin{align}
    \bar{n}^\mu = \frac{1}{\sqrt{2}}(\hat{t}^\mu + \hat{z}^\mu),
    \quad
    n^\mu = \frac{1}{\sqrt{2}}(\hat{t}^\mu - \hat{z}^\mu),
    \label{e.lc1}
\end{align} 
we can express the transverse metric $g_T^{\mu\nu}$ and the transverse antisymmetric tensor $\epsilon_T^{\mu\nu}$ with external momenta as
\begin{align}
    g_T^{\mu \nu}&=g^{\mu\nu}- \frac{P^\mu q^\nu+P^\nu q^\mu}{P\cdot q \left(1-\gamma_h^2\right)} +\frac{\gamma_h^2}{1-\gamma_h^2}\left(\frac{q^\mu q^\nu}{Q^2}+\frac{P^\mu P^\nu}{M^2}\right),\\
    \epsilon_T^{\mu\nu}&=\frac{1}{P\cdot q \sqrt{1-\gamma_h^2}}\epsilon^{\mu\nu\rho\sigma}q_\rho P_\sigma.
\end{align}
Furthermore, we set the lepton momentum in the $\hat{x}-\hat{z}$ plane and define the transverse spatial basis vectors as
\begin{align}
\hat{x}^\mu = \frac{ g_T^{\mu\nu} l_{1\nu}}{\sqrt{-g_T^{\mu\nu}l_{1\mu}l_{1\nu}}},
\quad
\hat{y}^\mu =\epsilon_T^{\mu\nu} \hat{x}_\nu.
\end{align}

In the coordinate system given by $\{ \hat{t}, \hat{x}, \hat{y}, \hat{z} \}$, one can explicitly write down the momenta as
\begin{align}
l_1^\mu&=\frac{Q}{2} \left(1, \sin{\theta}, 0, \cos{\theta}  \right),\\
l_2^\mu&=\frac{Q}{2} \left(1, -\sin{\theta}, 0, -\cos{\theta}  \right),\\
P^\mu&=\frac{z_h Q}{2} \left(1, 0, 0, -\sqrt{1-\gamma_h^2} \right),\\
q^\mu&=(Q,0,0,0),
\end{align}
where $\theta$ is the polar angle of the electron momentum $l_1$. It can be defined in Lorentz invariant form as
\begin{align}
\cos{\theta}=\frac{2y-1}{\sqrt{1-\gamma_h^2}},
\quad
\sin{\theta}=\sqrt{\frac{4y-4y^2-\gamma_h^2}{1-\gamma_h^2}},
\end{align}
where $y = (P \cdot l_1) / (P \cdot q)$. 

By construction, the hadron momentum $P$ has zero transverse component and the large component is $P^-$. So the longitudinal and transverse components of the spin tensors in Eqs.~\eqref{e.spin_s}--\eqref{e.spin_r} have the same meanings in this reference frame and coordinate system. In addition to $\hat{x}$ and $\hat{y}$, which can be used to project out the transverse components, we define a longitudinal basis vector,
\begin{align}
\hat{L}^\mu =-\frac{M}{2 P\cdot \bar{n}}\bar{n}^\mu+\frac{P\cdot \bar{n} }{M}n^\mu,
\end{align}
which satisfies the orthogonal relation $P\cdot \hat{L} = 0$. Then we redefine the spin components in Eqs.~\eqref{e.components1}, \eqref{e.components2}, and~\eqref{e.components3} as
\begin{align}
S_L &= S^\mu \hat{L}_\mu \label{e.SL},\\
S_T^x &= |S_T|\cos\phi_T =-S^\mu\hat{x}_\mu, \label{e.STx}\\
S_T^y &= |S_T|\sin\phi_T =-S^\mu\hat{y}_\mu,\\
S_{LL} &=T^{\mu \nu}\hat{L}_\mu \hat{L}_\nu,\\
S_{LT}^x &= |S_{LT}|\cos\phi_{LT} =-2 T^{\mu \nu}\hat{L}_\mu \hat{x}_\nu,\\
S_{LT}^y &= |S_{LT}|\sin\phi_{LT} =-2 T^{\mu \nu}\hat{L}_\mu \hat{y}_\nu,\\
S_{TT}^{xx} &= |S_{TT}|\cos 2\phi_{TT} =2 T^{ \mu \nu}\hat{x}_\mu \hat{x}_\nu + T^{\mu \nu}\hat{L}_\mu \hat{L}_\nu,\\
S_{TT}^{xy} &= |S_{TT}|\sin 2\phi_{TT} =2 T^{ \mu \nu}\hat{x}_\mu \hat{y}_\nu, \\
S_{LLL} &= R^{\mu \nu \rho}\hat{L}_\mu \hat{L}_\nu \hat{L}_\rho,\\
S_{LLT}^x &= |S_{LLT}|\cos\phi_{LLT} =-R^{\mu \nu \rho}\hat{L}_\mu \hat{L}_\nu \hat{x}_\rho,\\
S_{LLT}^y &= |S_{LLT}|\sin\phi_{LLT} =-R^{\mu \nu \rho}\hat{L}_\mu \hat{L}_\nu \hat{y}_\rho,\\
S_{LTT}^{xx} &= |S_{LTT}|\cos 2\phi_{LTT} =4R^{ \mu \nu \rho}\hat{L}_\mu \hat{x}_\nu \hat{x}_\rho + 2R^{\mu \nu \rho}\hat{L}_\mu \hat{L}_\nu \hat{L}_\rho,\\
S_{LTT}^{xy} &= |S_{LTT}|\sin 2\phi_{LTT} =4R^{\mu \nu \rho}\hat{L}_\mu \hat{x}_\nu \hat{y}_\rho,\\
S_{TTT}^{xxx} &= |S_{TTT}|\cos 3\phi_{TTT} =-4R^{ \mu \nu \rho}\hat{x}_\mu \hat{x}_\nu \hat{x}_\rho - 3 R^{\mu \nu \rho}\hat{L}_\mu \hat{L}_\nu \hat{x}_\rho,\\
S_{TTT}^{yxx} &= |S_{TTT}|\sin 3\phi_{TTT} =-4R^{\mu \nu \rho}\hat{y}_\mu \hat{x}_\nu \hat{x}_\rho - R^{\mu \nu \rho}\hat{L}_\mu \hat{L}_\nu \hat{y}_\rho. \label{e.STTTyxx}
\end{align}
Since $\hat{x}$, $\hat{y}$, and $\hat{L}$ are constructed with external momenta of the reaction, the quantities $S_L$, $|S_T|$, $\phi_T$, $S_{LL}$, $|S_{LT}|$, $\phi_{LT}$, $|S_{TT}|$, $\phi_{TT}$, $S_{LLL}$, $|S_{LLT}|$, $\phi_{LLT}$, $|S_{LTT}|$, $\phi_{LTT}$, $|S_{TTT}|$, and $\phi_{TTT}$ are Lorentz scalars, although the meaning of transverse or longitudinal direction is more clearly understood in this particular frame. The differential cross section can be expressed as
\begin{align}
P^{0} \frac{d \sigma}{d^{3} \bm{P}}=& \frac{\alpha^{2}}{ Q^{4}}
\left\{ \left(1+\cos^2{\theta}\right)F_U^T+ \left(1-\cos^2{\theta}\right)F_U^L+ |S_T|\sin{\phi_T}\sin{2\theta}F_T^{\sin{\phi_T}} \right. \nonumber\\ 
& + S_{LL} \left[\left(1+\cos^2{\theta}\right)F_{LL}^T + \left(1-\cos^2{\theta}\right)F_{LL}^L \right] \nonumber\\ 
&+|S_{LT}| \cos{\phi_{LT}}\sin{2\theta} F_{LT}^{\cos{\phi_{LT}}} + |S_{TT}| \cos{2\phi_{TT}} \sin^2{\theta} F_{TT}^{\cos{2\phi_{TT}}} \nonumber\\
&\left. + |S_{LLT}|\sin{\phi_{LLT}}\sin{2\theta} F_{LLT}^{\sin{\phi_{LLT}}} + |S_{LTT}| \sin{2\phi_{LTT}}\sin^2{\theta} F_{LTT}^{\sin{2\phi_{LTT}}} \right\} \label{e.dsigma2},
\end{align}
where the $F$'s, as functions of $z_h$ and $Q^2$, are structure functions, which are linear combinations of the $V$'s in Eq.~\eqref{e.wmunu}. The nine structure functions are defined according to their contributions to certain hadron polarization states as labeled by the subscripts and the azimuthal modulations as labeled by the superscripts. For those contributing flat distributions in the azimuthal angles, we add superscripts $T$ and $L$ to represent the transverse and longitudinal polarization states of the virtual photon,  which lead to $\left(1+\cos^2{\theta}\right)$ and $\left(1-\cos^2{\theta}\right)$ distributions in the polar angle, respectively. The two structure functions, $F_{LLT}^{\sin\phi_{LLT}}$ and $F_{LTT}^{\sin 2\phi_{LTT}}$, are newly introduced for the production of rank-3 tensor polarized hadrons from unpolarized $e^+e^-$ collisions, while the other ones also exist in the production of a spin-1 hadron~\cite{Chen:2016moq}.

Now we calculate the structure functions in the parton model. According to the factorization, we write the hadronic tensor,
\begin{align}
 W^{\mu \nu}(q ; P , S,  T,  R )=\frac{1}{2\pi} \int d^4k \operatorname{Tr}[\Delta(k ;P, S , T , R ) H^{\mu \nu}(k,q)],
\end{align}
as a convolution between the correlation function, in Eq.~\eqref{e.correlator}, and the hard factor
\begin{align}
H^{\mu \nu}(k,q)&=\gamma^\mu (\slashed{q}-\slashed{k})\gamma^\nu   (2\pi) \delta_+((q-k)^2),
\end{align}
where the subscript ``+'' of $\delta$ function indicates picking the positive energy solution.
With the collinear approximation, one can derive the the hadronic tensor at leading twist as 
\begin{align}
 W^{\mu \nu}(q ; P , S,  T,  R )&=\frac{2}{z} \mathrm{Tr}\left[ \Delta(z) \gamma^\mu \slashed{\bar{n}} \gamma^\nu \right], \label{e.wtrace}
\end{align}
where the $k_T$-integrated correlation function $\Delta(z)$ is defined in Eq.~\eqref{e.deltaz}. As has been derived in Eqs.~\eqref{e.deltaU}--\eqref{e.deltaLLT}, it is parametrized in terms of collinear FFs,
\begin{align}
\Delta(z)=& \frac{1}{4} \Big\{ D_1\left(z \right) \slashed{n} + G_{1L} \left(z \right) S_L \gamma_5 \slashed{n} + H_{1 T}\left(z \right) i \sigma_{\mu \nu} \gamma_{5} n^{\mu} S_{T}^{\nu} \nonumber\\
&+ D_{1LL}\left(z \right) S_{LL} \slashed{n} + H_{1 LT}\left(z \right)  \sigma_{\mu \nu} S_{LT}^{\mu} n^{\nu} \nonumber\\
&+ G_{1LLL}\left(z \right)S_{LLL} \gamma_5\slashed{n} + H_{1 LLT}\left(z \right) i \sigma_{\mu \nu} \gamma_{5} n^{\mu} S_{LLT}^{\nu} \Big\}.
\label{e.deltaz_ff}
\end{align}

In this process, the collinear FF to $\Omega$ does not couple to other long-distance functions. As a consequence of the helicity conservation, the chiral-odd FFs decouple from the cross section in the limit of massless quarks, and only the chiral-even FFs contribute. Besides, for the unpolarized collision, the leptonic tensor $L_{\mu\nu}$ is symmetric under the one-photon-exchange approximation, and thus we only need to consider the symmetric part of the hadronic tensor $W^{\mu\nu}$. After contracting the leptonic tensor and the hadronic tensor, we obtain the differential cross section in terms of collinear FFs as
\begin{align}
P ^{0} \frac{d \sigma}{d^{3} \bm{P} }=N_c\sum_q e_q^2 \frac{ \alpha^{2}}{ Q^{4}} \frac{1}{z_h} \left(1+\cos^2{\theta} \right) \left[ D_{1,q}(z_h) + D_{1LL,q}(z_h)S_{LL} \right],\label{e.dsigma3}
\end{align}
where the sum of ``$q$'' runs over all active quark and antiquark flavors with $e_q$ being the charge in unit of the positron charge and a subscript ``$q$'' is explicitly added to the FFs. As expected from the above analysis, only two terms, $D_1$ and $D_{1LL}$, contribute to the cross section. Comparing with general expression from kinematic analysis~\eqref{e.dsigma2}, we can write the structure functions in terms of collinear FFs at leading twist as
\begin{align}
F_U^T (z_h, Q^2) &= \frac{N_c}{z_h} \sum_q e_q^2 D_{1,q}(z_h,Q^2),\\
F_{LL}^T (z_h, Q^2) &= \frac{N_c}{z_h} \sum_q e_q^2 D_{1LL,q}(z_h,Q^2),
\end{align}
where $N_c = 3$ is the number of colors. Since none of the newly defined rank-3 tensor polarized FFs contributes, the unpolarized differential cross section of the inclusive production of a spin-3/2 hadron shares the same form as that of the inclusive production of a spin-1 hadron. Therefore, one cannot extract rank-3 tensor polarized FFs from the measurement of inclusive production of a spin-3/2 hadron in unpolarized $e^+e^-$ annihilation, or in other words, any nonzero result of the measurement of a spin-3/2 hadron in a rank-3 tensor polarization state will indicate high-twist or high-order effects.

\subsection{Semi-inclusive production of the $\Omega$ in unpolarized $e^+ e^-$ collisions}
\label{s.semi-inclusive}

\begin{figure}[ht]
    \centering 
    \includegraphics[width=0.35\textwidth]{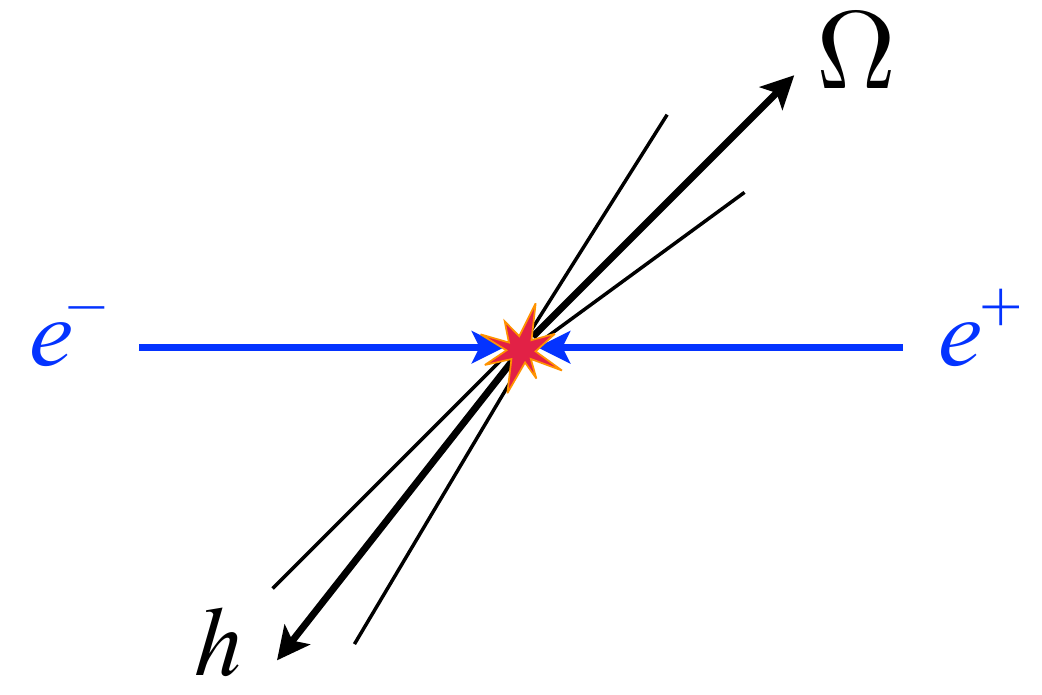}
    \hspace{1cm}
    \includegraphics[width=0.32\textwidth]{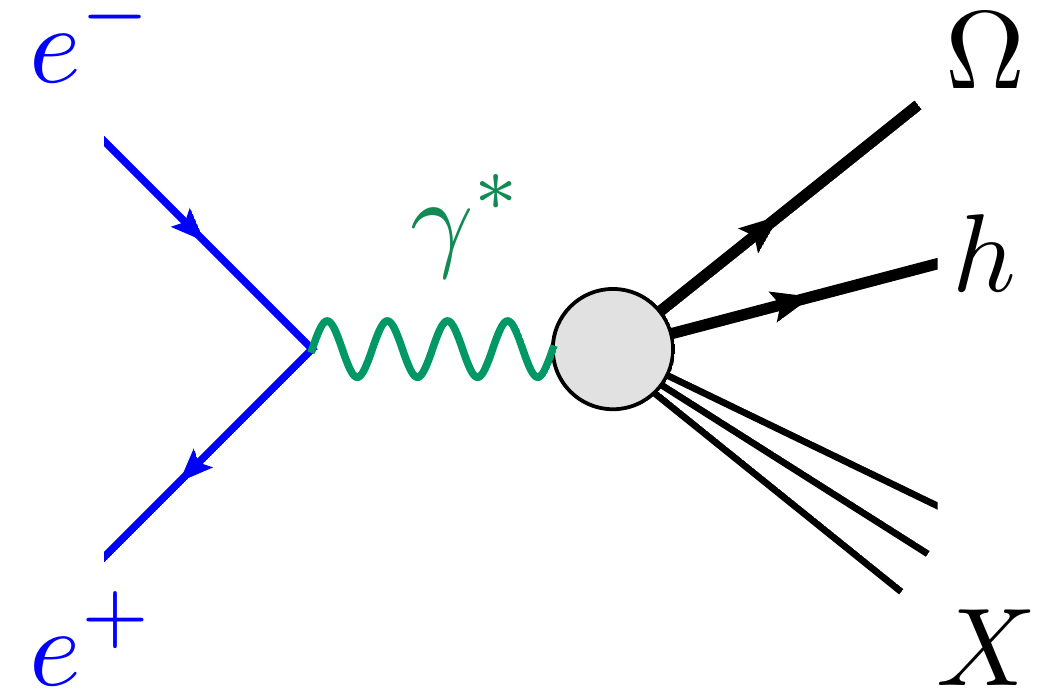}
    \caption{{\bf Left:} The illustration of the semi-inclusive production of $\Omega$ in $e^{+} e^{-}$ annihilation with the second observed hadron $h$ nearly back to the $\Omega$. {\bf Right:} The diagram of the process $e^+e^- \to \Omega h X$ under the one-photon-exchange approximation.}
\label{f.omegapiX}
\end{figure}

To access TMD FFs, we consider the process
\begin{align}
    e^- (l_1) + e^+ (l_2) \to \Omega (P_1) + h (P_2) + X (P_X),
\end{align}
as illustrated in Fig.~\ref{f.omegapiX}, where the variables in parentheses are the four momenta of the corresponding particles and for simplicity the spin of the second hadron $h$ is not taken into account. With one-photon-exchange approximation, the differential cross section can be expressed as
\begin{align}
 \frac{P_1^0 P_2^0 d \sigma}{d^{3} \bm{P}_1 d^{3} \bm{P}_2}=\frac{\alpha^{2}}{4 Q^{6}} L_{\mu \nu} W^{\mu \nu}, 
 \label{e.semidsigma}
\end{align}
where the leptonic tensor $L_{\mu\nu}$ has the same definition in Eq.~\eqref{e.lepton} and the hadronic tensor $W^{\mu\nu}$ is given by
\begin{align}
W^{\mu \nu}\left(q ; P_1, S, T, R; P_2\right)
&=
\frac{1}{(2 \pi)^4} \sum_X (2 \pi)^{4} \delta^{4}\left(q-P_{X}-P_1-P_2 \right)\nonumber\\ 
&\quad\times
\left\langle 0\left|J^{\mu}(0)\right| P_{X} ; P_1, S, T, R;P_2 \right\rangle  
\left\langle P_{X} ; P_1, S, T, R; P_2 \left|J^{\nu}(0)\right| 0\right\rangle,
\label{e.hadrontensor2}
\end{align}
where $S$, $T$, and $R$ represent the spin states of the first hadron, while the second hadron is unpolarized. The hadronic tensor satisfies the Hermiticity and the parity invariance relations,
\begin{align}
W^{* \mu \nu}\left( q ; P_1,S,T,R; P_2\right)
&=W^{\nu \mu}\left( q ; P_1,S,T,R; P_2\right), \\ 
W^{\mu \nu}\left( q ; P_1,S,T,R; P_2\right)
&=W_{\mu \nu}\left(q ; \bar{P}_1,-\bar{S},\bar{T},-\bar{R}; \bar{P}_2\right), 
\end{align}
and the gauge invariance requirement $q_\mu W^{\mu\nu} = W^{\mu\nu} q_\nu = 0$.

Following the same procedure in Sec.~\ref{s.inclusive}, one can, in principle, construct the basis tensors from the conserved vectors and tensors by exhausting all possible combinations. However, it is not efficient with increasing number of measured momenta. Instead of directly constructing all basis tensors, we have a systematic procedure for the semi-inclusive process, where the hadronic tensor depends on three external momenta. Because the four vectors $P_1^\mu$, $P_2^\mu$, $q^\mu$, and $\epsilon^{\mu q P_1 P_2}$ form complete basis of the spacetime, we can first construct the {\it basic Lorentz tensors},
\begin{align}
t_{U}^{  \mu \nu} &= \left\{ \widetilde{g}^{\mu \nu}, \widetilde{P}_{1}^{\mu} \widetilde{P}_{1}^{\nu}, \widetilde{P}_{2}^{\mu} \widetilde{P}_{2}^{\nu}, \widetilde{P}_{1}^{\{\mu} \widetilde{P}_{2}^{\nu \}} \right\},\\
t_{U}^{ \mathcal{P}, \mu \nu} &= \left\{\widetilde{P}_{1}^{\{\mu} \epsilon^{\nu\} q P_{1} P_{2}}, \widetilde{P}_{2}^{\{\mu} \epsilon^{\nu\} q P_{1} P_{2}}\right\},
\end{align}
which are grouped into two sets $t_U^{\mu\nu}$ and $t_U^{{\cal P},\mu\nu}$, corresponding to parity conserving and parity nonconserving ones, respectively.
The conserved vectors $\widetilde{P}_{1}$ and $\widetilde{P}_{2}$ are defined in a similar way to Eq.~\eqref{e.Ptilde}. The unpolarized basis tensors are directly given by those in $t_U^{\mu\nu}$. One may also include those in $t_U^{{\cal P}\mu\nu}$ if parity-violating channels, e.g., the exchange of a $Z$ boson, are taken into account. Since all the spin components can essentially be expressed into scalars or pseudoscalars, as explicitly written in Eqs.~\eqref{e.SL}--\eqref{e.STTTyxx}, the polarized basis tensors can be obtained by multiplying the basic Lorentz tensors by a spin-dependent scalar or pseudoscalar. Considering the parity conserving case, we multiply the basic Lorentz tensors in $t_U^{\mu\nu}$ by a spin-dependent scalar and multiply those in $t_U^{{\cal P},\mu\nu}$ by a spin-dependent pseudoscalar. Then the eight vector polarized basis tensors, 16 rank-2 tensor polarized basis tensors, and 20 rank-3 tensor polarized basis tensors are constructed from those in $t_U^{\mu\nu}$ and $t_U^{{\cal P},\mu\nu}$ as
\begin{align}
t_{V}^{  \mu \nu}&=\left\{ 
\epsilon^{S q P_{1} P_{2}} \right\} t_{U}^{  \mu \nu}, 
\left\{S \cdot q , S \cdot P_{2} \right\} t_{U}^{ \mathcal{P}, \mu \nu},\\
t_{T}^{  \mu \nu}&=\left\{T^{P_2 P_2}, T^{P_{2} q}, T^{q q}\right\} t_{U}^{  \mu \nu}, \left\{\epsilon^{T^{P_2}P_1 P_2 q }, \epsilon^{T^{q}P_1 P_2 q }\right\}t_{U}^{ \mathcal{P}, \mu \nu},\\
t_{R}^{  \mu \nu}&=
\left\{\epsilon^{R^{P_2 P_2}P_1 P_2 q },\epsilon^{R^{P_2 q}P_1 P_2 q }, \epsilon^{R^{q q}P_1 P_2 q }\right\}t_{U}^{ \mu \nu},
\left\{R^{P_2 P_2 P_2},R^{q q q},R^{P_2 P_2 q},R^{P_2 q q}\right\}t_{U}^{ \mathcal{P}, \mu \nu}.
\end{align}
With the 48 basis tensors above, we can expand the hadronic tensor $W^{\mu\nu}$ as
\begin{align}
W^{  \mu \nu}=\sum_{i=1}^{4} V_{U, i}  t_{U, i}^{  \mu \nu} +
\sum_{i=1}^{8} V_{V, i}  t_{V, i}^{  \mu \nu} + \sum_{i=1}^{16} V_{T, i}  t_{T, i}^{  \mu \nu}+\sum_{i=1}^{20} V_{R, i}  t_{R, i}^{  \mu \nu}, \label{e.totaltensor}
\end{align}
where the coefficients $V$'s are scalar functions of $q^2$, $P_1\cdot q$, $P_2\cdot q$, and $P_1 \cdot P_2$. We should note that this procedure is invalid for the inclusive process because the number of momenta in the hadronic tensor is not enough to form a complete basis of the spacetime.

Contracting the hadronic tensor with the leptonic tensor, one can derive the differential cross section in Eq.~\eqref{e.semidsigma}, and it is common to introduce some dimensionless variables to replace the scalar products of momenta,
\begin{align}
z_{h1}&=\frac{2P_1 \cdot q}{Q^2}, 
\quad 
z_{h2}=\frac{2P_2 \cdot q}{Q^2},
\quad 
\xi=\frac{2P_1\cdot P_2}{Q^2},  
\nonumber\\
y_1&=\frac{l_1\cdot P_1}{q\cdot P_1}, 
\quad 
y_2=\frac{l_1\cdot P_2}{q\cdot P_2},
\quad
\gamma_{h1}=\frac{2M_1}{z_{h1} Q}, 
\quad 
\gamma_{h2}=\frac{2M_2}{z_{h2} Q},
\end{align}
where $M_1$ and $M_2$ are the masses of the two hadrons, respectively.

\begin{figure}[ht]
    \centering 
    \includegraphics[width=0.5\textwidth]{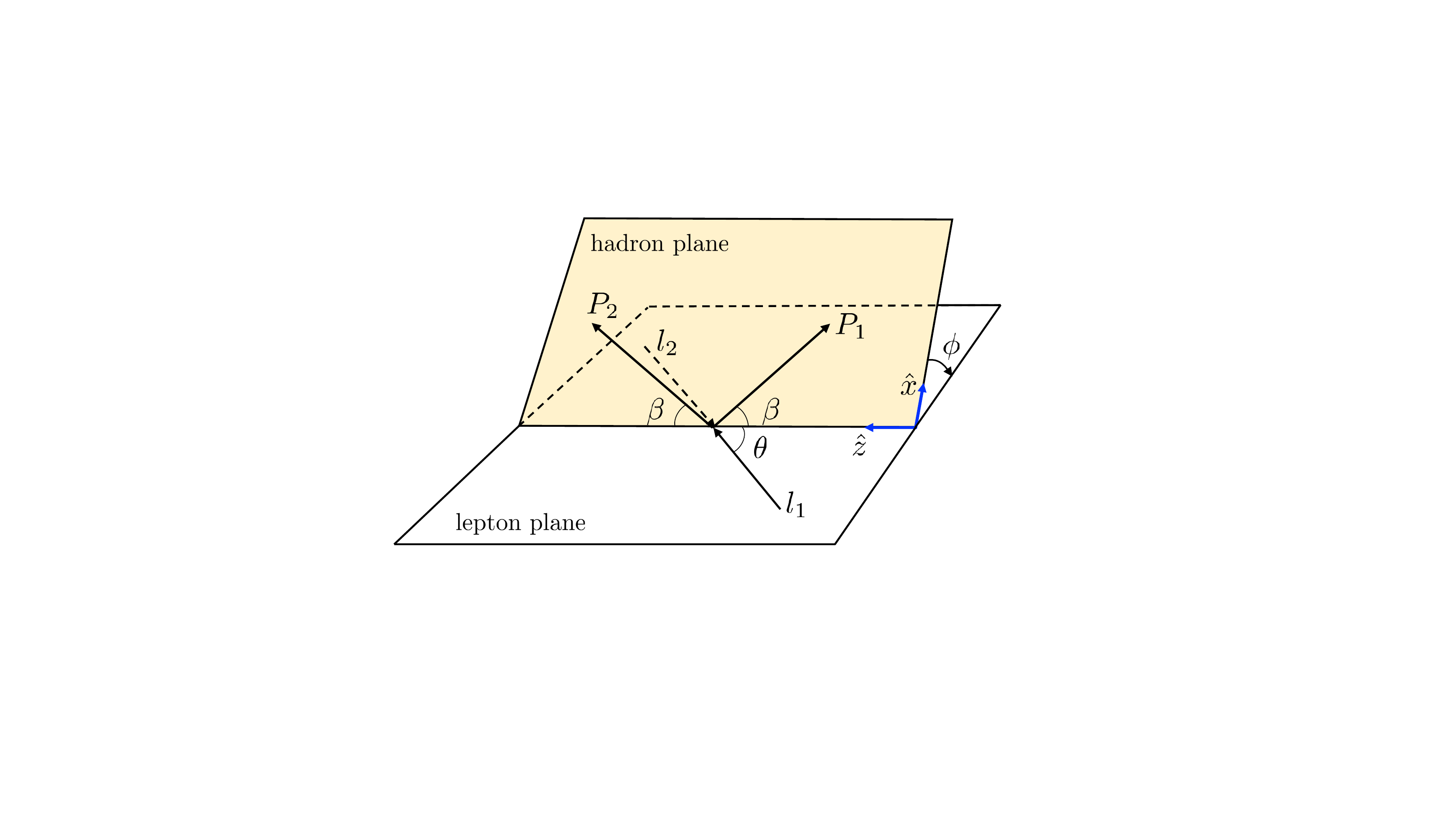}
    \caption{The Collins-Soper frame for $e^{+} e^{-}\rightarrow \Omega h X$.}
\label{f.CS}
\end{figure}

To separate contributions from different structure functions, it is in practice convenient to specify some reference frame and coordinate system to define angular distributions. For this process, there are two commonly used frames.
One is the Collins-Soper (CS) frame~\cite{Collins:1977iv}, illustrated in Fig.~\ref{f.CS}. It is defined as the center-of-mass frame of the leptons with the longitudinal spatial axis chosen as the exterior angle bisector of the momenta of the two hadrons in the final state. In this frame, one can write down the components of the momenta as
\begin{align}
l_{1}^{\mu} &=\frac{Q}{2}(1, \sin \theta \cos \phi, \sin \theta \sin \phi, \cos \theta),\\
l_{2}^{\mu} &=\frac{Q}{2}(1,-\sin \theta \cos \phi,-\sin \theta \sin \phi,-\cos \theta), \\
P_{1}^{\mu} & = \frac{z_{h1}Q}{2}\left(1, \sqrt{1-\gamma_{h1}^2}\sin\beta ,0, -\sqrt{1-\gamma_{h1}^2}\cos\beta \right), \\
P_{2}^{\mu} & = \frac{z_{h2}Q}{2}\left(1, \sqrt{1-\gamma_{h2}^2}\sin\beta ,0, \sqrt{1-\gamma_{h2}^2}\cos\beta \right),\\
q^{\mu} & =(Q,0,0,0),
\end{align}
where the angles $\beta$, $\theta$, and $\phi$, as illustrated in Fig.~\ref{f.CS}, can also be defined in Lorentz invariant forms,
\begin{align}
\cos{2\beta}&=\frac{2\xi-z_{h1} z_{h2}}{z_{h1} z_{h2} \sqrt{\left(1-\gamma_{h1}^2\right)}\sqrt{\left(1-\gamma_{h2}^2\right)}}
\approx \frac{2\xi-z_{h1} z_{h2}}{z_{h1} z_{h2} }, \\
\cos{\theta}&=\frac{1-2y_2}{2\sqrt{1-\gamma_{h2}^2}\cos{\beta}}-\frac{1-2y_1}{2\sqrt{1-\gamma_{h1}^2}\cos{\beta}}
\approx \frac{y_1-y_2}{\cos{\beta}},\label{e.cstheta}\\  
\cos{\phi}&=\frac{1-2y_2}{2\sqrt{1-\gamma_{h2}^2}\sin{\beta}\sin{\theta}} + \frac{1-2y_1}{2\sqrt{1-\gamma_{h1}^2}\sin{\beta}\sin{\theta}}
\approx \frac{1-y_1-y_2}{\sin{\beta\sin{\theta}}}, \label{e.csphi}
\end{align}
where ``$\approx$'' represents the approximation of neglecting the hadron masses although the masses are kept in this calculation. 

\begin{figure}[ht]
    \centering
    \includegraphics[width=0.5\textwidth]{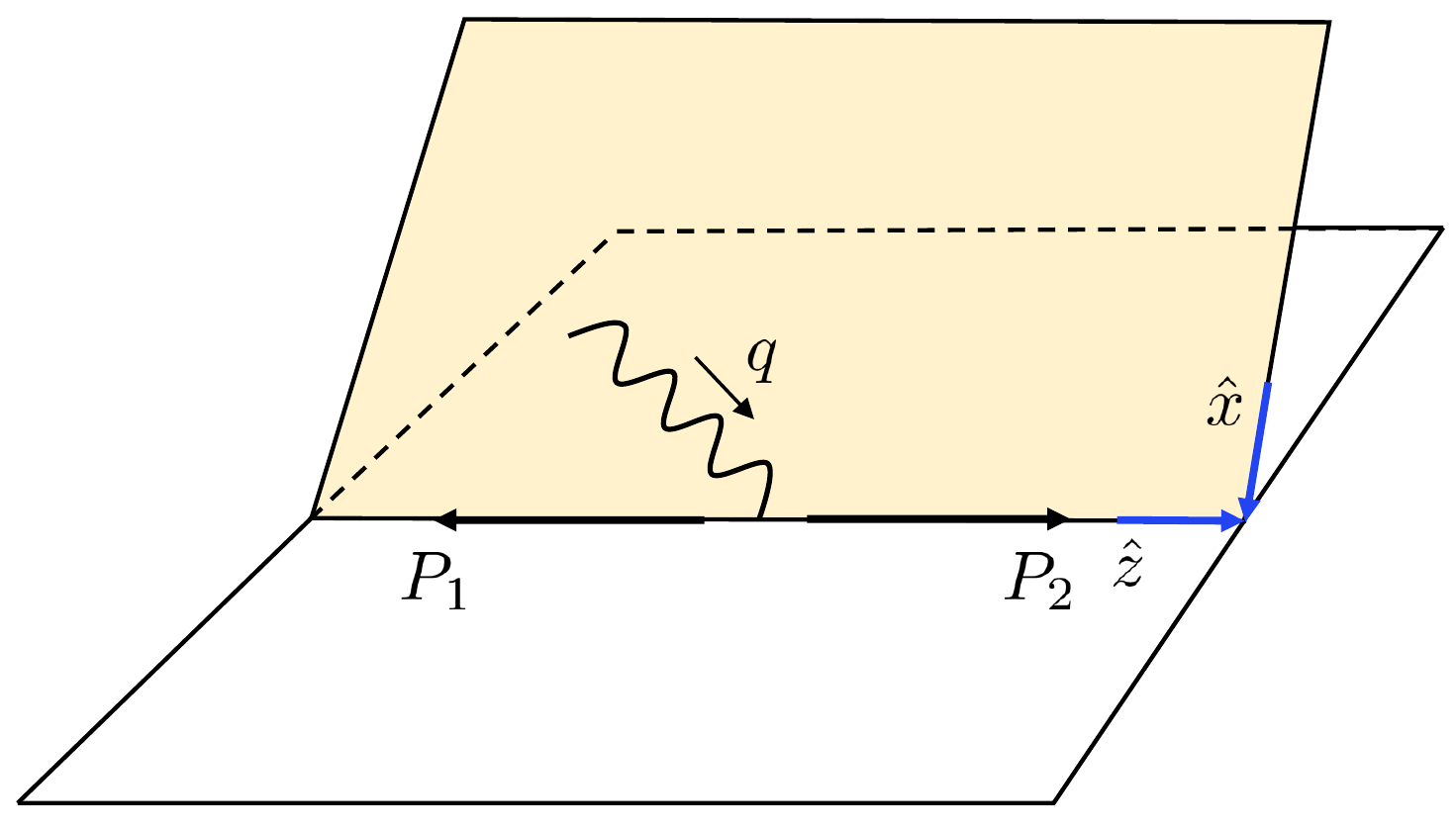}
    \caption{The two-hadron center-of-mass frame for $e^{+} e^{-}\rightarrow \Omega h X$.}
\label{f.cm}
\end{figure}

The other frame is the two-hadron back-to-back frame, which can be chosen as the center-of-mass frame of the two hadrons in the final state, referred to as the c.m. frame. As illustrated in Fig.~\ref{f.cm}, we construct the coordinate system by choosing the temporal basis vector as
\begin{align}
\hat{t}^\mu&=\frac{P_1^\mu+P_2^\mu}{\sqrt{M_1^2 + M_2^2 + 2P_1\cdot P_2}}, \label{e.that}
\end{align}
and the longitudinal spatial basis vector as
\begin{align}
    \hat{z}^\mu&=\frac{P_2^\mu \left( M_1^2 + P_1\cdot P_2\right) - P_1^\mu \left( M_2^2 + P_1\cdot P_2\right)}{\sqrt{\left(P_1 \cdot P_2\right)^2-M_1^2 M_2^2} \sqrt{M_1^2 + M_2^2 + 2P_1\cdot P_2}}.\label{e.zhat}
\end{align}
The light-cone basis vectors are given by $\bar{n}^\mu = (\hat{t}^\mu + \hat{z}^\mu) / \sqrt{2}$ and ${n}^\mu = (\hat{t}^\mu - \hat{z}^\mu) / \sqrt{2}$, which are different from those in Eq.~\eqref{e.lc1}. Correspondingly, the transverse metric tensor $g_T^{\mu\nu}$ and the transverse antisymmetric tensor $\epsilon_T^{\mu\nu}$ are given by
\begin{align}
g^{\mu \nu}_T&=g^{\mu \nu}-\frac{\left( P_1\cdot  P_2\right) \left( P_1^{\mu }  P_2^{\nu }+ P_1^{\nu }  P_2^{\mu }\right)}{\left(P_1 \cdot P_2\right)^2-M_1^2 M_2^2} 
+ \frac{M_1^2  P_2^{\mu }  P_2^{\nu }+M_2^2  P_1^{\mu }  P_1^{\nu }}{\left(P_1 \cdot P_2\right)^2-M_1^2 M_2^2}\nonumber\\
&\approx g^{\mu \nu}- \frac{  P_1^{\mu }  P_2^{\nu }+ P_1^{\nu }  P_2^{\mu }}{P_1\cdot P_2},\\
\epsilon_T^{\mu\nu}&=\epsilon^{\mu\nu\rho\sigma} \frac{P_{2\rho} P_{1\sigma}}{\sqrt{\left(P_1 \cdot P_2\right)^2-M_1^2 M_2^2}}\nonumber\\
&\approx \epsilon^{\mu\nu\rho\sigma} \frac{P_{2\rho} P_{1\sigma}}{P_1\cdot P_2}.
\end{align}
By construction, the hadrons have zero transverse momenta and the large momentum components are $P_1^-$ and $P_2^+$ according to this choice of $\hat{t}$ and $\hat{z}$. Whereas, the virtual photon has transverse momentum in this frame. We choose the transverse spatial basis vectors as
\begin{align}
    \hat{x}^\mu&=\frac{g_T^{\mu\nu}q_\nu}{\sqrt{-g_T^{\mu\nu} q_\mu q_\nu}},
    \label{e.cm_x}\\
    \hat{y}^\mu&=\epsilon_T^{\mu \nu }\hat{x}_\nu,
    \label{e.cm_y}
\end{align}
and thus the virtual photon is in the $\hat{x}-\hat{z}$ plane.

In this calculation, we find the CS frame is more convenient to describe the angular distributions of the produced hadrons, but the spin components are easier to be defined in the c.m. frame. Therefore, taking advantage of the both frames, we will use $\theta$ and $\phi$ introduced in the CS frame and the spin components in the c.m. frame, $S_L$, $|S_T|$, $\phi_T$, $S_{LL}$, $|S_{LT}|$, $\phi_{LT}$, $|S_{TT}|$, $\phi_{TT}$, $S_{LLL}$, $|S_{LLT}|$, $\phi_{LLT}$, $|S_{LTT}|$, $\phi_{LTT}$, $|S_{TTT}|$, and $\phi_{TTT}$, which have the same expressions as Eqs.~\eqref{e.SL}--\eqref{e.STTTyxx} but with the basis vectors replaced by Eqs.~\eqref{e.cm_x} and~\eqref{e.cm_y}. Since all these variables are defined in Lorentz invariant forms, one can in principle evaluate them in arbitrary frame, although the meanings of the variables are clearer in certain frames.

After contracting the hadronic tensor and the leptonic tensor, we can express the differential cross section in terms of 48 structure functions according to the angular distributions and the spin states of the produced hadrons,
\begin{align}
&\frac{P_{1}^{0} P_{2}^{0} d \sigma}{d^{3} \bm{P}_{1} d^{3} \bm{P}_{2}}=\frac{\alpha^{2}}{4 Q^{4}} \times \nonumber\\
&\Bigg\{\left[\left(1+\cos ^{2} \theta\right) F_{U, U}^T+\left(1-\cos ^{2} \theta\right) F_{U, U}^L+(\sin 2 \theta \cos \phi) F_{U, U}^{\cos \phi}+\left(\sin ^{2} \theta \cos 2 \phi\right) F_{U, U}^{\cos 2 \phi}\right] \nonumber \\
&+S_{L}\left[\left(\sin ^{2} \theta \sin 2 \phi\right) F_{L, U}^{\sin 2 \phi}+(\sin 2 \theta \sin \phi) F_{L, U}^{\sin \phi}\right] \nonumber\\
&+\left|{S}_{T}\right|\left[\operatorname { s i n } \phi _ {T} \left(\left(1+\cos ^{2} \theta\right) F_{T, U}^T+\left(1-\cos ^{2} \theta\right) F_{T, U}^L+(\sin 2 \theta \cos \phi) F_{T, U}^{\cos \phi}\right.\right.\nonumber\\
&\left.\left.+\left(\sin ^{2} \theta \cos 2 \phi\right) F_{T, U}^{\cos 2 \phi}\right)+\cos \phi_{T}\left(\left(\sin ^{2} \theta \sin 2 \phi\right) F_{T, U}^{\sin 2 \phi}+(\sin 2 \theta \sin \phi) F_{T, U}^{\sin \phi}\right)\right]\nonumber\\
&+S_{LL}\left[\left(1+\cos ^{2} \theta\right) F_{LL,U}^T+\left(1-\cos ^{2} \theta\right) F_{LL,U}^L+(\sin 2 \theta \cos \phi) F_{LL,U}^{\cos \phi}+\left(\sin ^{2} \theta \cos 2 \phi\right) F_{LL,U}^{\cos 2 \phi}\right]\nonumber\\
&+\left|S_{LT}\right|\left[\operatorname { cos } \phi _ { LT } \left(\left(1+\cos ^{2} \theta\right) F_{LT,U}^T+\left(1-\cos ^{2} \theta\right) F_{LT,U}^L+(\sin 2 \theta \cos \phi) F_{LT,U}^{\cos \phi}\right.\right.\nonumber\\
&\left.\left.+\left(\sin ^{2} \theta \cos 2 \phi\right) F_{LT,U}^{\cos 2 \phi}\right)+\sin \phi_{LT}\left(\left(\sin ^{2} \theta \sin 2 \phi\right) F_{LT,U}^{\sin 2 \phi}+(\sin 2 \theta \sin \phi) F_{LT,U}^{\sin \phi}\right)\right]\nonumber\\
&+\left|S_{TT}\right|\left[\operatorname { cos } 2\phi _ { TT } \left(\left(1+\cos ^{2} \theta\right) F_{TT,U}^T+\left(1-\cos ^{2} \theta\right) F_{TT,U}^L+(\sin 2 \theta \cos \phi) F_{TT,U}^{\cos \phi}\right.\right.\nonumber\\
&\left.\left.+\left(\sin ^{2} \theta \cos 2 \phi\right) F_{TT,U}^{\cos 2 \phi}\right)+\sin 2\phi_{TT}\left(\left(\sin ^{2} \theta \sin 2 \phi\right) F_{TT,U}^{\sin 2 \phi}+(\sin 2 \theta \sin \phi) F_{TT,U}^{\sin \phi}\right)\right]\nonumber\\
&+S_{LLL}\left[\left(\sin ^{2} \theta \sin 2 \phi\right) F_{LLL,U}^{\sin 2 \phi}+(\sin 2 \theta \sin \phi) F_{LLL,U}^{\sin \phi}\right]\nonumber\\
&+\left|S_{LLT}\right|\left[\operatorname { sin } \phi _ { LLT } \left(\left(1+\cos ^{2} \theta\right) F_{LLT,U}^T+\left(1-\cos ^{2} \theta\right) F_{LLT,U}^L+(\sin 2 \theta \cos \phi) F_{LLT,U}^{\cos \phi}\right.\right.\nonumber\\
&\left.\left.+\left(\sin ^{2} \theta \cos 2 \phi\right) F_{LLT,U}^{\cos 2 \phi}\right)+\cos \phi_{LLT}\left(\left(\sin ^{2} \theta \sin 2 \phi\right) F_{LLT,U}^{\sin 2 \phi}+(\sin 2 \theta \sin \phi) F_{LLT,U}^{\sin \phi}\right)\right]\nonumber\\
&+\left|S_{LTT}\right|\left[\operatorname { sin } 2\phi _ {LTT } \left(\left(1+\cos ^{2} \theta\right) F_{LTT,U}^T+\left(1-\cos ^{2} \theta\right) F_{LTT,U}^L+(\sin 2 \theta \cos \phi) F_{LTT,U}^{\cos \phi}\right.\right.\nonumber\\
&\left.\left.+\left(\sin ^{2} \theta \cos 2 \phi\right) F_{LTT,U}^{\cos 2 \phi}\right)+\cos 2\phi_{LTT}\left(\left(\sin ^{2} \theta \sin 2 \phi\right) F_{LTT,U}^{\sin 2 \phi}+(\sin 2 \theta \sin \phi) F_{LTT,U}^{\sin \phi}\right)\right]\nonumber\\
&+\left|S_{TTT}\right|\left[\operatorname { sin } 3\phi _ {TTT } \left(\left(1+\cos ^{2} \theta\right) F_{TTT,U}^T+\left(1-\cos ^{2} \theta\right) F_{TTT,U}^L+(\sin 2 \theta \cos \phi) F_{TTT,U}^{\cos \phi}\right.\right.\nonumber\\
&\left.\left. +\left(\sin ^{2} \theta \cos 2 \phi\right) F_{TTT,U}^{\cos 2 \phi}\right)+\cos 3\phi_{TTT}\left(\left(\sin ^{2} \theta \sin 2 \phi\right) F_{TTT,U}^{\sin 2 \phi}+(\sin 2 \theta \sin \phi) F_{TTT,U}^{\sin \phi}\right)\right]\Bigg\}, \label{e.semicrosssection}
\end{align}
where the two subscripts for each structure function represent the polarization states of the two hadrons, $\Omega$ and $h$, respectively, though the spin of the second hadron is not taken into account and thus effectively unpolarized. The superscripts have the same meanings as those in the single hadron inclusive production process, either labeling the azimuthal modulation or representing the virtual photon polarization. 

The $F$'s, as scalar functions of $z_{h1}$, $z_{h2}$, $Q^2$, and ${\bm q}_T^2 \equiv -g_T^{\mu\nu} q_\mu q_\nu$, are linear combinations of the $V$'s in Eq.~\eqref{e.totaltensor}. Among them, the 20 rank-3 tensor polarized ones are newly defined and only exist when one of the detected two hadrons has spin $s\ge 3/2$, while the four unpolarized ones, the eight vector polarized ones, and the 16 rank-2 tensor polarized ones also exist if one of the observed hadrons is spin-1.

Now we calculate the structure functions in the parton model and only focus on the kinematic region in which the produced two hadrons are nearly back to back. It corresponds to the region ${\bm q}_T^2 \ll Q^2$, where one can apply the TMD factorization. At the leading order, we express the hadronic tensor $W^{\mu\nu}$ as the production of a quark-antiquark pair convoluting with two correlation functions,
\begin{align}
W^{\mu \nu}=  N_c z_1 z_2 \sum_q e_q^2 \int d^2\bm{k}_{1T}  d^2\bm{k}_{2T} \delta^{(2)} (\bm{k}_{1T}+\bm{k}_{2T}-\bm{q}_{T}) {\rm Tr}[\Delta^{\Omega/q}\gamma^\mu \Delta^{h/\bar{q}}\gamma^\nu] , \label{e.hadrontensorffs}
\end{align}
where the transverse momenta ${\bm k}_{1T}$, ${\bm k}_{2T}$, and ${\bm q}_T$ are defined in the c.m. frame. Since the two hadrons have zero transverse momentum components in this frame, the ${\bm k}_{1T}$ and ${\bm k}_{2T}$ directly correspond to the quark transverse momentum in the definition of the correlation function, while the longitudinal direction is flipped for the second hadron in comparison to Eq.~\eqref{e.unintcorrrelator}. Hence, the light-cone longitudinal momentum fractions are given by $z_1 = k_1^- / P_1^-$ and $z_2 = k_2^+ / P_2^+$. The superscripts ``$\Omega/q$'' and ``$h / \bar{q}$'' of the $\Delta$'s indicate the relevant quark and hadron, and will be suppressed in the rest of the paper if no confusion is caused. The sum on ``$q$" runs over all active flavors of quark and antiquark.

Alternative to the complete parametrization in Eqs.~\eqref{e.tmdff_u}--\eqref{e.tmdff_ttt}, we write the $\Delta$'s in a concise form,
\begin{align}
\Delta(z_1, k_{1T}) &= 
\frac{1}{4} \left \{ \Delta^{\left[\gamma^{-}\right]}(z_1, k_{1T}) \gamma^{+}
-\Delta^{\left[\gamma^{-} \gamma_{5}\right]}(z_1, k_{1T}) \gamma^{+} \gamma_{5}
+\Delta^{\left[i \sigma^{i-} \gamma_{5}\right]}(z_1, k_{1T}) i \sigma_{-i} \gamma_{5} \right\},\label{e.Delta}\\
\Delta(z_{2}, k_{2T}) &=
\frac{1}{4} \left \{ \Delta^{\left[\gamma^{+}\right]}(z_{2}, k_{2T}) \gamma^{-}
-\Delta^{\left[\gamma^{+} \gamma_{5}\right]}(z_{2}, k_{2T}) \gamma^{-} \gamma_{5}
+\Delta^{\left[i \sigma^{j+} \gamma_{5}\right]}(z_{2}, k_{2T}) i \sigma_{+j} \gamma_{5} \right\},\label{e.Deltabar}
\end{align}
where we use the notation
    $
    \Delta^{[\Gamma]}\left(z, k_{T}\right) \equiv
    {\rm Tr}\left[\Delta\left(z, k_{T}\right) \Gamma\right]
    $
with $\Gamma=\gamma^-$, $\gamma^-\gamma_5$, and $i\sigma^{i-}\gamma_5$ for the hadron $\Omega$ and $\Gamma=\gamma^+$, $\gamma^+\gamma_5$, and $i\sigma^{i+}\gamma_5$ for the hadron $h$. A complete parametrization of $\Delta^{[\Gamma]}(z,k_T)$'s  in terms of TMD FFs are provided in Appendix~\ref{a.FFs}. Since $h$ is unpolarized, there are only two leading-twist TMD FFs, which are defined as
\begin{align}
\Delta^{[\gamma^+]} (z_{2}, k_{2T}) &= D_1(z_{2}, k_{2T}^2),\\
\Delta^{[i \sigma^{i+} \gamma_5]} (z_{2}, k_{2T}) &=- \frac{\epsilon_T^{ij} k_{2Tj}}{M_2} H_1^{\perp}(z_{2}, k_{2T}^2).
\end{align}
Substituting the FFs in Eq.~\eqref{e.hadrontensorffs} with Eqs.~\eqref{e.Delta} and~\eqref{e.Deltabar}, one can easily find that only two terms
\begin{align*}
    {\rm Tr}\Big[\Delta^{\left[\gamma^{-}\right]}(z_1, k_{1T}) \gamma^{+}
    \gamma^\mu
    \Delta^{\left[\gamma^{+}\right]}(z_{2}, k_{2T}) \gamma^{-}
    \gamma^\nu
    \Big],\\
    {\rm Tr}\Big[\Delta^{\left[i \sigma^{i-} \gamma_{5}\right]}(z_1, k_{1T}) i \sigma_{-i} \gamma_{5}
    \gamma^\mu
    \Delta^{\left[i \sigma^{j+} \gamma_{5}\right]}(z_{2}, k_{2T}) i \sigma_{+j} \gamma_{5}
    \gamma^\nu
    \Big],
\end{align*}
contribute to the symmetric part of $W^{\mu\nu}$, because the chiral-odd TMD FFs in $\Delta^{[i\sigma^{i-}\gamma_5]}$ must couple to another chiral-odd function for the helicity conservation of massless quarks. Then the differential cross section at the leading-twist approximation is
\begin{align}
\frac{P_{1}^{0} P_{2}^{0} d \sigma}{d^{3} \bm{P}_{1} d^{3} \bm{P}_{2}}&=\frac{\alpha^{2}}{4 Q^{4}} \frac{1}{4}N_c z_{1} z_{2} \sum_{q} e_{q}^{2} \int d^{2} \bm{k}_{1 T} d^{2} \bm{k}_{2 T} \delta^{(2)} (\bm{k}_{1 T}+\bm{k}_{2 T}-\bm{q}_{T} ) \nonumber\\
&\times\left\{ \left(1+\cos ^{2} \theta\right)\Delta^{\left[\gamma^{-}\right]}(z_1, k_{1T}) \Delta^{\left[\gamma^{+}\right]}(z_2, k_{2T})
+\left[ \sin ^{2} \theta \cos 2 \phi\left(\delta^{i 1} \delta^{j 1}-\delta^{i 2} \delta^{j 2}\right)\right.\right. \nonumber\\
&\left.\left.+\sin^{2}\theta \sin 2\phi\left(\delta^{i 1} \delta^{j 2}+\delta^{i 2} \delta^{j 1}\right)\right] \Delta^{\left[i \sigma^{i-} \gamma_{5}\right]}(z_1, k_{1T}) \Delta^{\left[i \sigma^{j+} \gamma_{5}\right]}(z_2, k_{2T})
\right\}. \label{e.crosssectionffs}
\end{align}

Comparing Eq.~\eqref{e.crosssectionffs} with Eq.~\eqref{e.semicrosssection}, we can obtain the structure functions, which are convolutions of two TMD FFs. For conciseness, we introduce the transverse momentum convolution notation
\begin{align}
&{\cal C}\left[w_a(k_{ 1 T}, k_{ 2 T}) D(z_1, k_{1T}^2) D(z_2, k_{2T}^2)\right] 
\nonumber\\
&\equiv \frac{1}{4} N_{c} z_1 z_2  \sum_{q} e_{q}^{2} 
\int d^{2} \bm{k}_{ 1 T} d^{2} \bm{k}_{ 2 T} \delta^{(2)}(\bm{k}_{ 1 T}+\bm{k}_{ 2 T}-\bm{q}_{T})
w_a(k_{ 1 T}, k_{ 2 T}) D_q(z_1, k_{1T}^{2}) D_{\bar q}(z_2, k_{2T}^{2}),
\end{align}
where $D_q(z_ 1, k_{1T}^2)$ is a TMD FF for the first hadron $\Omega$, $D_{\bar q}(z_2,k_{2T}^2)$ is a TMD FF for the second hadron $h$, and $w_a(k_{ 1 T}, k_{ 2 T})$, with $a=1,\cdots,10$, is one of the dimensionless scalar functions as provided below,
\begin{align}
&w_{1} = -\frac{\hat{q}_{\scriptscriptstyle T} \cdot k_{\scriptscriptstyle 1T}}{M_{1}}, \quad
w_2 = -\frac{\hat{q}_{\scriptscriptstyle T} \cdot k_{\scriptscriptstyle 2 T}}{M_{2}} ,
\quad w_3 = \frac{2(\hat{q}_{\scriptscriptstyle T} \cdot k_{\scriptscriptstyle 1 T}) (\hat{q}_{\scriptscriptstyle T} \cdot k_{\scriptscriptstyle 2 T})+k_{\scriptscriptstyle 1 T} \cdot k_{\scriptscriptstyle2 T}}{M_{1} M_{2}},\nonumber \\
&w_4 = \frac{k_{\scriptscriptstyle 1T}^{ij} \hat{q}_{\scriptscriptstyle Ti} k_{{\scriptscriptstyle 2Tj}} + 2k_{\scriptscriptstyle 1T}^{ij} \hat{q}_{\scriptscriptstyle Ti} \hat{q}_{\scriptscriptstyle Tj} (\hat{q}_{\scriptscriptstyle T}\cdot k_{\scriptscriptstyle 2T})}{M_1^2 M_2},
\quad w_5 = \frac{2k_{\scriptscriptstyle 1T}^{ij}\hat{q}_{\scriptscriptstyle Ti} \hat{q}_{\scriptscriptstyle Tj}}{M_1^2},\nonumber\\
&w_6 = \frac{2\left[k_{\scriptscriptstyle 1T}^{ijl}\hat{q}_{\scriptscriptstyle Ti} \hat{q}_{\scriptscriptstyle Tj} k_{\scriptscriptstyle 2Tl} + 2k_{\scriptscriptstyle 1T}^{ijl}\hat{q}_{\scriptscriptstyle Ti} \hat{q}_{\scriptscriptstyle Tj} \hat{q}_{\scriptscriptstyle Tl} (k_{\scriptscriptstyle 2T}\cdot \hat{q}_{\scriptscriptstyle T})\right]}{M_1^3 M_2}, 
\quad w_7= -\frac{k_{\scriptscriptstyle 1 T} \cdot k_{\scriptscriptstyle 2 T}}{M_{1} M_{2}},
\quad w_8 = \frac{4k_{\scriptscriptstyle 1T}^{ijl}\hat{q}_{\scriptscriptstyle Ti} \hat{q}_{\scriptscriptstyle Tj} \hat{q}_{\scriptscriptstyle Tl}}{M_1^3},\nonumber\\
&w_9 = \frac{4\left[k_{\scriptscriptstyle 1T}^{ijlm}\hat{q}_{\scriptscriptstyle Ti} \hat{q}_{\scriptscriptstyle Tj} \hat{q}_{\scriptscriptstyle Tl} k_{\scriptscriptstyle 2Tm} +2k_{\scriptscriptstyle 1T}^{ijlm} \hat{q}_{\scriptscriptstyle Ti} \hat{q}_{\scriptscriptstyle Tj} \hat{q}_{\scriptscriptstyle Tl} \hat{q}_{\scriptscriptstyle Tm} (k_{\scriptscriptstyle 2T}\cdot \hat{q}_{\scriptscriptstyle T})\right]}{M_1^4 M_2},
\quad w_{10} = \frac{2k_{\scriptscriptstyle 1T}^{ij} \hat{q}_{\scriptscriptstyle Ti} k_{\scriptscriptstyle 2Tj}}{M_1^2 M_2},
\end{align}
where $\hat{q}_T^\mu \equiv g_T^{\mu\nu} q_\nu / \sqrt{{\bm q}_T^2}$ is the direction of the virtual photon transverse momentum in the c.m. frame.

At leading twist, 24 structure functions have nontrivial expressions. Two of the nonvanishing structure functions are for unpolarized hadron state,
\begin{align}
F_{U, U}^{T} &= {\cal C}\left[D_{1}(z_1, k_{1T}^2) D_{1}(z_2, k_{2T}^2)\right], 
\label{e.Fuu1}\\
F_{U, U}^{\cos 2 \phi} &= {\cal C}\left[w_3(k_{ 1 T}, k_{ 2 T}) H_{1}^{\perp}(z_1, k_{1T}^2) H_{1}^{\perp}(z_2, k_{2T}^2)\right],\label{e.Fuu2}
\end{align}
where the $\cos{2\phi}$ azimuthal modulation term $F_{U, U}^{\cos 2 \phi}$ is widely used to extract the Collins function $H_{1}^{\perp}(z,k_T^2)$~\cite{Belle:2005dmx,Belle:2008fdv,Belle:2019nve,BaBar:2013jdt}.
Four of the nonvanishing structure functions are for vector polarized hadron states, including one for $S_L$ dependence,
\begin{align}
F_{L, U}^{\sin 2 \phi}&=-{\cal C}\left[w_3(k_{ 1 T}, k_{ 2 T}) H_{1 L}^{\perp}(z_1, k_{1T}^2) H_{1}^{\perp}(z_2, k_{2T}^2)\right], 
\end{align} 
and three for $S_T$ dependence,
\begin{align}
F_{T, U}^{T}&={\cal C}\left[w_{1}(k_{ 1 T}, k_{ 2 T}) D_{1 T}^{\perp}(z_1, k_{1T}^2) D_{1}(z_2, k_{2T}^2)\right], \\
F_{T, U}^{\sin \left(2 \phi+\phi_{T}\right)}&=\frac{1}{2}\left(F_{T, U}^{\cos 2 \phi}+F_{T, U}^{\sin 2 \phi}\right)={\cal C}\left[w_4(k_{ 1 T}, k_{ 2 T}) H_{1 T}^{\perp}(z_1, k_{1T}^2) H_{1}^{\perp}(z_2, k_{2T}^2)\right], \\
F_{T, U}^{\sin \left(2 \phi-\phi_{T}\right)}&=\frac{1}{2}\left(F_{T, U}^{\sin 2 \phi}-F_{T, U}^{\cos 2 \phi}\right)=-{\cal C}\left[w_2(k_{ 1 T}, k_{ 2 T}) H_{1T}(z_1, k_{1T}^2) H_{1}^{\perp}(z_2, k_{2T}^2)\right].
\end{align}
Eight of the nonvanishing structure functions are for rank-2 tensor polarized hadron states, including two for $S_{LL}$ dependence,
\begin{align}
F_{LL,U}^{T}&={\cal C}\left[D_{1LL}(z_1, k_{1T}^2) D_{1}(z_2, k_{2T}^2)\right],
\\
F_{LL,U}^{\cos{2\phi}}&=-{\cal C}[w_3(k_{ 1 T}, k_{ 2 T}) H_{1LL}^\perp(z_1, k_{1T}^2) H_{1}^{\perp}(z_2, k_{2T}^2)],
\end{align}
three for $S_{LT}$ dependence,
\begin{align}
F_{LT,U}^{T}&={\cal C}\left[w_{1}(k_{ 1 T}, k_{ 2 T}) D_{1 LT}^\perp(z_1, k_{1T}^2) D_{1}(z_2, k_{2T}^2)\right], \\
F_{LT,U}^{\sin \left(2 \phi+\phi_{LT}\right)}&=\frac{1}{2}\left(F_{LT,U}^{\cos 2 \phi}+F_{LT,U}^{\sin 2 \phi}\right)=-{\cal C}\left[w_2(k_{ 1 T}, k_{ 2 T}) H_{1 LT}(z_1, k_{1T}^2) H_{1}^{\perp}(z_2, k_{2T}^2)\right],\\
F_{LT,U}^{\sin \left(2 \phi-\phi_{LT}\right)}&=\frac{1}{2}\left(F_{LT,U}^{\sin 2 \phi}-F_{LT,U}^{\cos 2 \phi}\right)=-{\cal C}\left[{w}_{4}(k_{ 1 T}, k_{ 2 T}) H_{1 LT}^\perp(z_1, k_{1T}^2) H_{1}^{\perp}(z_2, k_{2T}^2)\right],
\end{align}
and three for $S_{TT}$ dependence,
\begin{align}
F_{TT,U}^{T}&={\cal C}\left[ w_5(k_{ 1 T}, k_{ 2 T}) D_{1TT}^\perp(z_1, k_{1T}^2) D_{1}(z_2, k_{2T}^2)\right],\\
F_{TT,U}^{\sin \left(2 \phi+2\phi_{TT}\right)}&=\frac{1}{2}\left(F_{TT,U}^{\cos 2 \phi}+F_{TT,U}^{\sin 2 \phi}\right)={\cal C}\left[{w}_{7}(k_{ 1 T}, k_{ 2 T}) H_{1 TT}^\perp(z_1, k_{1T}^2)  H_{1}^{\perp}(z_2, k_{2T}^2)\right],\\
F_{TT,U}^{\sin \left(2 \phi-2\phi_{TT}\right)}&=\frac{1}{2}\left(F_{TT,U}^{\sin 2 \phi}-F_{TT,U}^{\cos 2 \phi}\right)={\cal C}\left[w_6(k_{ 1 T}, k_{ 2 T}) H_{1 TT}^{\perp\perp}(z_1, k_{1T}^2) H_{1}^{\perp}(z_2, k_{2T}^2)\right].
\end{align}
Ten of the nonvanishing structure functions are for rank-3 tensor polarized hadron states, including one for $S_{LLL}$ dependence,
\begin{align}
F_{LLL,U}^{\sin 2 \phi}=-{\cal C}\left[w_3(k_{ 1 T}, k_{ 2 T}) H_{1 LLL}^{\perp}(z_1, k_{1T}^2) H_{1}^{\perp}(z_2, k_{2T}^2)\right],
\end{align}
three for $S_{LLT}$ dependence,
\begin{align}
F_{LLT,U}^{T}&={\cal C}\left[w_{1}(k_{ 1 T}, k_{ 2 T}) D_{1 LLT}^{\perp}(z_1, k_{1T}^2) D_{1}(z_2, k_{2T}^2)\right], \\
F_{LLT,U}^{\sin \left(2 \phi+\phi_{LLT}\right)}&=\frac{1}{2}\left(F_{LLT,U}^{\cos 2 \phi}+F_{LLT,U}^{\sin 2 \phi}\right)={\cal C}\left[{w}_{4}(k_{ 1 T}, k_{ 2 T}) H_{1 LLT}^\perp(z_1, k_{1T}^2) H_{1}^{\perp}(z_2, k_{2T}^2)\right],\\
F_{LLT,U}^{\sin \left(2 \phi-\phi_{LLT}\right)}&=\frac{1}{2}\left(F_{LLT,U}^{\sin 2 \phi}-F_{LLT,U}^{\cos 2 \phi}\right)=-{\cal C}\left[w_2(k_{ 1 T}, k_{ 2 T}) H_{1 LLT}(z_1, k_{1T}^2) H_{1}^{\perp}(z_2, k_{2T}^2)\right],
\end{align} 
three for $S_{LTT}$ dependence,
\begin{align}
F_{LTT,U}^{T}&=-{\cal C}\left[ w_5(k_{ 1 T}, k_{ 2 T}) D_{1LTT}^\perp(z_1, k_{1T}^2) D_{1}(z_2, k_{2T}^2)\right],\\
F_{LTT,U}^{\sin \left(2 \phi+2\phi_{LTT}\right)}&=\frac{1}{2}\left(F_{LTT,U}^{\cos 2 \phi}+F_{LTT,U}^{\sin 2 \phi}\right) =-{\cal C}\left[w_6(k_{ 1 T}, k_{ 2 T}) H_{1 LTT}^{\perp \perp}(z_1, k_{1T}^2) H_{1}^{\perp}(z_2, k_{2T}^2)\right], \\
F_{LTT,U}^{\sin \left(2 \phi-2\phi_{LTT}\right)}&=\frac{1}{2}\left(F_{LTT,U}^{\sin 2 \phi}-F_{LTT,U}^{\cos 2 \phi}\right)={\cal C}\left[w_7(k_{ 1 T}, k_{ 2 T}) H_{1 LTT}^\perp(z_1, k_{1T}^2) H_{1}^{\perp}(z_2, k_{2T}^2)\right],
\end{align} 
and three for $S_{TTT}$ dependence,
\begin{align}
F_{TTT,U}^{T}&=-{\cal C}\left[ w_8(k_{ 1 T}, k_{ 2 T}) D_{1TTT}^\perp(z_1, k_{1T}^2) D_{1}(z_2, k_{2T}^2)\right],\\
F_{TTT,U}^{\sin \left(2 \phi+3\phi_{TTT}\right)}&=\frac{1}{2}\left(F_{TTT,U}^{\cos 2 \phi}+F_{TTT,U}^{\sin 2 \phi}\right) =-{\cal C}\left[w_9(k_{ 1 T}, k_{ 2 T}) H_{1 TTT}^{\perp \perp}(z_1, k_{1T}^2) H_{1}^{\perp}(z_2, k_{2T}^2)\right],\\
F_{TTT,U}^{\sin \left(2 \phi-3\phi_{TTT}\right)}&=\frac{1}{2}\left(F_{TTT,U}^{\sin
2 \phi}-F_{TTT,U}^{\cos 2 \phi}\right) =-{\cal C}\left[w_{10}(k_{ 1 T}, k_{ 2 T}) H_{1 TTT}^\perp(z_1, k_{1T}^2) H_{1}^{\perp}(z_2, k_{2T}^2)\right].\label{e.FTTTU}
\end{align}
The measurement of these 24 structure functions, particularly the ten for rank-3 tensor polarized states, can be utilized to study the TMD FFs for a spin-3/2 hadron, although the other 24 structure functions in the unpolarized differential cross section only arise at high twist or high order.

\section{Summary and outlook} 
\label{s.summary}

In this paper, we have studied the inclusive and semi-inclusive production of spin-3/2 hadrons in $e^+e^-$ annihilation. 

We describe the spin state of a hadron with the spin density matrix $\rho$. For a spin-3/2 hadron, its spin density matrix contains 16 independent components, including one for unpolarized state, three for vector polarized states, five for rank-2 tensor polarized states, and seven for rank-3 tensor polarized states. By imposing orthogonal relations, we are able to use the basis matrices as spin projectors to extract the spin components of the hadron state. 

Considering existing $e^+e^-$ collision experiments, such as Belle II, BaBar, and BESIII, we only focus on unpolarized collisions and the parity violating effect via a $Z$-boson exchange is neglected. Through kinematic analyses, we derive the differential cross section of inclusive production of a spin-3/2 hadron in unpolarized $e^+e^-$ collisions in terms of nine structure functions, among which $F_{LLT}^{\sin\phi_{LLT}}$ and $F_{LTT}^{\sin 2\phi_{LTT}}$ are newly defined and only exist when spin of the produced hadron no less than $3/2$. For the process with two identified hadrons, we only consider the spin of one observed hadron. Taking advantage of one additional observed momentum, we can simplify the kinematic analysis of the hadronic tensor with a systematic procedure, in which all basis tensors are constructed from spin-independent basic Lorentz tensors multiplied by a spin-dependent scalar or pseudoscalar. Then the differential cross section are expressed in terms of 48 structure functions in accordance to the angular distributions and spin states of the produced hadrons. Among the 48 structure functions, 20 are newly defined, corresponding to the production of rank-3 tensor polarized hadron state, while the other 28 also exist for the production of a spin-1 hadron and an unpolarized hadron.

Applying the factorized formalism, we express the cross sections of these reactions as convolutions between the production of quarks at a hard scale and some long-distance functions describing the hadronization process. At leading twist, we provide a complete parametrization of the long-distance function, i.e., the quark-quark correlation function, in terms of TMD FFs, including two for the unpolarized hadron state, six for the vector polarized hadron states, ten for rank-2 tensor polarized hadron states, and 14 for rank-3 tensor polarized states. At the collinear limit only seven of them are nonvanishing, including $D_1(z)$ for the unpolarized hadron state, $G_{1L}(z)$ and $H_{1T}(z)$ for the vector polarized hadron states, $D_{1LL}(z)$ and $H_{1LT}(z)$ for rank-2 tensor polarized hadron states, and $G_{1LLL}(z)$ and $H_{1LLT}(z)$ for rank-3 tensor polarized hadron states. 

For the inclusive process, the reaction is not sensitive to quark transverse momentum and one applies the collinear factorization. We perform a leading order calculation in the parton model and find that none of the FFs for rank-3 tensor polarized hadron states contributes to the cross section. Therefore, any observation of a rank-3 tensor polarized hadron state, as well as vector polarized hadron state, from such unpolarized collision will imply high-twist or high-order effects.

For the semi-inclusive process, we consider the kinematic region ${\bm q}_T^2 \ll Q^2$ and apply the TMD factorization formalism. Through a leading order calculation, we find half of the structure functions vanish at leading twist, but ten of the leading-twist nontrivial structure functions are contributions from rank-3 tensor polarized hadron states, and therefore can be utilized to study the rank-3 tensor polarized TMD FFs for spin-3/2 hadrons.

The hadron production in high energy $e^+e^-$ annihilation, as the cleanest process to study FFs, has been widely used to extract unpolarized FF $D_1(z)$ and its TMD counterpart $D_1(z,k_T^2)$. The azimuthal asymmetry data of two hadron productions from Belle, BaBar, and BESIII have been combined with SIDIS data from HERMES, COMPASS, and JLab to extract the Collins function $H_{1}^{\perp}(x,k_T^2)$. Furthermore, the recent measurement of $\Lambda$ polarization by Belle has stimulated extensive studies of FFs to polarized hadron states. These results encourage us to explore more possibilities for the study of FFs and the role of spin in the hadronization process at $e^+e^-$ machines. The FFs to $\Omega$, a spin-3/2 baryon, may provide an opportunity, not only because it has more spin states and thus more observables, but also because it has unique sensitivity to strange quarks. 

Recently, the BESIII Collaboration measured the polarization of the $\Omega$ from $\psi(3686) \rightarrow \Omega^- \bar{\Omega}^+$ via a chain of weak decays~$\Omega^- \rightarrow K^- \Lambda$ and $\Lambda \rightarrow p \pi^-$~\cite{BESIII:2020lkm}. Although these data cannot be directly utilized to study FFs, the developed techniques to analyze the spin state of the produced $\Omega$ can be applied in other processes with $\Omega$ produced from parton fragmentation. The Belle II experiment with 40 times higher luminosity than the Belle experiment is expected to produce enough $\Omega$ events for the spin state analysis. Therefore, the analysis of the spin state of $\Omega$ and the extraction of the rank-3 tensor polarized FFs are in principle feasible with future data from the Belle II experiment.

One can also consider the spin of both hadrons in the semi-inclusive process, which will require much more statistics in experiment for the analysis of both spin states, which may be possible with the Belle II experiment. The FFs to $\Omega$ obtained in $e^+e^-$ annihilation can also be applied in the SIDIS process and have advantages to study the strange sea distributions in the nucleon.

Therefore, the study of TMD FFs to spin-3/2 hadrons will provide new opportunities for understanding the role of spin in the hadronization process. The semi-inclusive data of the $\Omega$ production at existing $e^+e^-$ collision experiments can be utilized to explore the new FFs and more efforts are expected from both theoretical and experimental sides.

\section*{Acknowledgements}
We thank Zhiqing Liu and Jinlong Zhang for useful discussions.
This work was supported by the National Natural Science Foundation of China (Approvals No. 12175117, No. 11890713, and No. 11890700) and Shandong Province Natural Science Foundation Grant No. ZR2020MA098.

\newpage

\appendix

\section{PROBABILITY INTERPRETATIONS OF THE SPIN COMPONENTS} 
\label{a.SDM}

A spin-3/2 hadron has four independent spin states and the corresponding spin density matrix $\rho$ is a $4\times4$ matrix, which in the $s_z$ representation can be explicitly written as
\begin{equation}
\rho=\frac{1}{4}\left(
\begin{array}{cccc}
\rho_{\frac{3}{2} \frac{3}{2}} & \rho_{\frac{3}{2} \frac{1}{2}} & \rho_{\frac{3}{2} -\frac{1}{2}} & \rho_{\frac{3}{2} -\frac{3}{2}} \\
\rho_{\frac{1}{2} \frac{3}{2}} & \rho_{\frac{1}{2} \frac{1}{2}} & \rho_{\frac{1}{2} -\frac{1}{2}} & \rho_{\frac{1}{2} -\frac{3}{2}} \\
\rho_{-\frac{1}{2} \frac{3}{2}} & \rho_{-\frac{1}{2} \frac{1}{2}} & \rho_{-\frac{1}{2} -\frac{1}{2}} & \rho_{-\frac{1}{2} -\frac{3}{2}}\\
\rho_{-\frac{3}{2} \frac{3}{2}} & \rho_{-\frac{3}{2} \frac{1}{2}} & \rho_{-\frac{3}{2} -\frac{1}{2}} & \rho_{-\frac{3}{2} -\frac{3}{2}}
\end{array}
\right).\label{e.densitymatrix}
\end{equation}
Comparing with the decomposition in Eq.~\eqref{e.density} and the spin vector and tensors, \eqref{e.spinvector}, \eqref{e.tensor1}, and \eqref{e.tensor2}, one can match the matrix elements with spin components as
\begin{align}
\rho_{\frac{3}{2} \frac{3}{2}}&=1+\frac{6}{5}S_{L}+S_{LL}+\frac{2}{3} S_{LLL}, \\
\rho_{\frac{1}{2} \frac{1}{2}}&=1+\frac{2}{5}S_{L}-S_{LL}- 2S_{LLL}, \\
\rho_{-\frac{1}{2} -\frac{1}{2}}&=1-\frac{2}{5}S_{L}-S_{LL}+ 2S_{LLL}, \\
\rho_{-\frac{3}{2} -\frac{3}{2}}&=1-\frac{6}{5}S_{L}+S_{LL}-\frac{2}{3} S_{LLL}, \\
\rho_{\frac{3}{2} \frac{1}{2}}&=\frac{2\sqrt{3}}{5}(S_T^x-iS_T^y)+ \frac{\sqrt{3}}{3}(S_{LT}^x- iS_{LT}^y)+\frac{2\sqrt{3}}{3}(S_{LLT}^x-iS_{LLT}^y),\\
\rho_{\frac{1}{2} \frac{3}{2}}&=\frac{2\sqrt{3}}{5}(S_T^x+iS_T^y)+ \frac{\sqrt{3}}{3}(S_{LT}^x+ iS_{LT}^y)+\frac{2\sqrt{3}}{3}(S_{LLT}^x+iS_{LLT}^y),\\
\rho_{\frac{1}{2} -\frac{1}{2}}&=\frac{4}{5}(S_T^x-iS_T^y)- 2(S_{LLT}^x-iS_{LLT}^y),\\
\rho_{-\frac{1}{2} \frac{1}{2} }&=\frac{4}{5}(S_T^x+iS_T^y)- 2(S_{LLT}^x+iS_{LLT}^y),\\
\rho_{-\frac{1}{2} -\frac{3}{2}}&=\frac{2\sqrt{3}}{5}(S_T^x-iS_T^y)- \frac{\sqrt{3}}{3}(S_{LT}^x- iS_{LT}^y)+\frac{2\sqrt{3}}{3}(S_{LLT}^x-iS_{LLT}^y),\\
\rho_{-\frac{3}{2} -\frac{1}{2} }&=\frac{2\sqrt{3}}{5}(S_T^x+iS_T^y)- \frac{\sqrt{3}}{3}(S_{LT}^x+ iS_{LT}^y)+\frac{2\sqrt{3}}{3}(S_{LLT}^x+iS_{LLT}^y),\\
\rho_{\frac{3}{2} -\frac{1}{2}}&=\frac{\sqrt{3}}{3}(S_{TT}^{xx}- iS_{TT}^{xy})+ \frac{\sqrt{3}}{3}(S_{LTT}^{xx}-iS_{LTT}^{xy}),\\
\rho_{-\frac{1}{2} \frac{3}{2} }&=\frac{\sqrt{3}}{3}(S_{TT}^{xx}+ iS_{TT}^{xy})+ \frac{\sqrt{3}}{3}(S_{LTT}^{xx}+iS_{LTT}^{xy}),\\
\rho_{\frac{1}{2} -\frac{3}{2}}&=\frac{\sqrt{3}}{3}(S_{TT}^{xx}-iS_{TT}^{xy})-\frac{\sqrt{3}}{3}(S_{LTT}^{xx}-iS_{LTT}^{xy}),\\
\rho_{\frac{1}{2} -\frac{3}{2}}&=\frac{\sqrt{3}}{3}(S_{TT}^{xx}+iS_{TT}^{xy})-\frac{\sqrt{3}}{3}(S_{LTT}^{xx}+iS_{LTT}^{xy}),\\
\rho_{\frac{3}{2} -\frac{3}{2}}&=\frac{2}{3}(S_{TTT}^{xxx}+iS_{TTT}^{yxx}),\\
\rho_{-\frac{3}{2} \frac{3}{2} }&=\frac{2}{3}(S_{TTT}^{xxx}-iS_{TTT}^{yxx}).
\end{align}

To analyze the probability interpretation of each spin component, it is convenient to choose the hadron rest frame, while the general situation for a moving hadron can be obtained via a Lorentz boost.  By measuring the spin along a particular direction
$\hat{\bm n} = (\sin\theta \cos\phi, \sin\theta \sin\phi, \cos\theta)$ specified by the polar angle $\theta$ and the azimuthal angle $\phi$, one will obtain an eigenstate of the spin operator,
\begin{align}
\Sigma^{i} \hat{n}_{i}=\Sigma^{x} \sin \theta \cos \phi+\Sigma^{y} \sin \theta \sin \phi+\Sigma^{z} \cos \theta.
\end{align}
Following the common convention, we represent the eigenstate by $\left|m_{(\theta,\phi)}\right>$, where $m = -3/2$, $-1/2$, $1/2$, and $3/2$ is the magnetic quantum number along the direction $(\theta, \phi)$. The probability of finding the specified eigenstate can be evaluated from the spin density matrix $\rho$ as
\begin{align}
P\left(m_{(\theta, \phi)}\right)=\operatorname{Tr}\left[\rho|m_{(\theta, \phi)}\rangle\langle m_{(\theta, \phi)}|\right].
\end{align}
Then all the spin components can be expressed in terms of the $P\left(m_{(\theta, \phi)}\right)$'s. The three spin components of the spin vector $S^i$, defined in Eq.~\eqref{e.components1}, are given by
\begin{align}
S_{L}=&\frac{3}{2}\left[P\left(\frac{3}{2}_{(0,0)}\right)-P\left(-\frac{3}{2}_{(0,0)}\right)\right]+\frac{1}{2}\left[P\left(\frac{1}{2}_{(0,0)}\right)-P\left(-\frac{1}{2}_{(0,0)}\right)\right], \\
S_{T}^{x}=&\frac{3}{2}\left[P\left(\frac{3}{2}_{(\frac{\pi}{2}, 0)}\right)-P\left(-\frac{3}{2}_{(\frac{\pi}{2}, 0)}\right)\right] +\frac{1}{2}\left[P\left(\frac{1}{2}_{(\frac{\pi}{2}, 0)}\right)-P\left(-\frac{1}{2}_{(\frac{\pi}{2}, 0)}\right)\right],\\
S_{T}^{y}=&\frac{3}{2}\left[P\left(\frac{3}{2}_{(\frac{\pi}{2}, \frac{\pi}{2})}\right)-P\left(-\frac{3}{2}_{(\frac{\pi}{2}, \frac{\pi}{2})}\right)\right]+\frac{1}{2}\left[P\left(\frac{1}{2}_{(\frac{\pi}{2}, \frac{\pi}{2})}\right)-P\left(-\frac{1}{2}_{(\frac{\pi}{2}, \frac{\pi}{2})}\right)\right]. 
\end{align}
The five spin components of the rank-2 spin tensor $T^{ij}$, defined in Eq.~\eqref{e.components2}, are given by
\begin{align}
S_{L L}=&\left[P\left(\frac{3}{2}_{(0,0)}\right)+P\left(-\frac{3}{2}_{(0,0)}\right)\right]-\left[P\left(\frac{1}{2}_{(0,0)}\right)+P\left(-\frac{1}{2}_{(0,0)}\right)\right], \\
S_{L T}^{x}=& 2\left\{\left[P\left(\frac{3}{2}_{(\frac{\pi}{4}, 0)}\right)+P\left(-\frac{3}{2}_{(\frac{\pi}{4}, 0)}\right)\right]-\left[P\left(\frac{3}{2}_{(-\frac{\pi}{4}, 0)}\right)+P\left(-\frac{3}{2}_{(-\frac{\pi}{4}, 0)}\right)\right]\right\}, \\
S_{L T}^{y}=& 2\left\{\left[P\left(\frac{3}{2}_{(\frac{\pi}{4}, \frac{\pi}{2})}\right)+P\left(-\frac{3}{2}_{(\frac{\pi}{4}, \frac{\pi}{2})}\right)\right]-\left[P\left(\frac{3}{2}_{(-\frac{\pi}{4}, \frac{\pi}{2})}\right)+P\left(-\frac{3}{2}_{(-\frac{\pi}{4}, \frac{\pi}{2})}\right)\right]\right\}, \\
S_{T T}^{x x}=& 2\left\{\left[P\left(\frac{3}{2}_{(\frac{\pi}{2}, 0)}\right)+P\left(-\frac{3}{2}_{(\frac{\pi}{2}, 0)}\right)\right]-\left[P\left(\frac{3}{2}_{(\frac{\pi}{2}, \frac{\pi}{2})}\right)+P\left(-\frac{3}{2}_{(\frac{\pi}{2}, \frac{\pi}{2})}\right)\right]\right\} ,\\
S_{T T}^{x y}=& 2\left\{\left[P\left(\frac{3}{2}_{(\frac{\pi}{2}, \frac{\pi}{4})}\right)+P\left(-\frac{3}{2}_{(\frac{\pi}{2}, \frac{\pi}{4})}\right)\right]-\left[P\left(\frac{3}{2}_{(\frac{\pi}{2}, -\frac{\pi}{4})}\right)+P\left(-\frac{3}{2}_{(\frac{\pi}{2}, -\frac{\pi}{4})}\right)\right]\right\}. 
\end{align}
The seven spin components of the rank-3 spin tensor $R^{ijk}$, defined in Eq.~\eqref{e.components3}, are given by
\begin{align}
S_{L L L}=& \frac{3}{10}\left[P\left(\frac{3}{2}_{(0,0)}\right)-P\left(-\frac{3}{2}_{(0,0)}\right)\right]-\frac{9}{10}\left[P\left(\frac{1}{2}_{(0,0)}\right)-P\left(-\frac{1}{2}_{(0,0)}\right)\right] ,\\
S_{L L T}^{x}=& -\frac{1}{60}\left\{129\left[P\left(\frac{3}{2}_{(\frac{\pi}{2}, 0)}\right)-P\left(-\frac{3}{2}_{(\frac{\pi}{2}, 0)}\right)\right]+23\left[P\left(\frac{1}{2}_{(\frac{\pi}{2}, 0)}\right)-P\left(-\frac{1}{2}_{(\frac{\pi}{2}, 0)}\right)\right]\right\} \nonumber\\
&+\frac{\sqrt{2}}{24}\left\{27\left[P\left(\frac{3}{2}_{(\frac{\pi}{4}, 0)}\right)-P\left(-\frac{3}{2}_{(\frac{\pi}{4}, 0)}\right)\right]+\left[P\left(\frac{1}{2}_{(\frac{\pi}{4}, 0)}\right)-P\left(-\frac{1}{2}_{(\frac{\pi}{4}, 0)}\right)\right]\right\} \nonumber\\
&+\frac{\sqrt{2}}{24}\left\{27\left[P\left(\frac{3}{2}_{(-\frac{\pi}{4}, 0)}\right)-P\left(-\frac{3}{2}_{(-\frac{\pi}{4}, 0)}\right)\right]+\left[P\left(\frac{1}{2}_{(-\frac{\pi}{4}, 0)}\right)-P\left(-\frac{1}{2}_{(-\frac{\pi}{4}, 0)}\right)\right]\right\},\\
S_{L L T}^{y}=& -\frac{1}{60}\left\{129\left[P\left(\frac{3}{2}_{(\frac{\pi}{2}, \frac{\pi}{2})}\right)-P\left(-\frac{3}{2}_{(\frac{\pi}{2}, \frac{\pi}{2})}\right)\right]+23\left[P\left(\frac{1}{2}_{(\frac{\pi}{2}, \frac{\pi}{2})}\right)-P\left(-\frac{1}{2}_{(\frac{\pi}{2}, \frac{\pi}{2})}\right)\right]\right\} \nonumber\\
&+\frac{\sqrt{2}}{24}\left\{27\left[P\left(\frac{3}{2}_{(\frac{\pi}{4}, \frac{\pi}{2})}\right)-P\left(-\frac{3}{2}_{(\frac{\pi}{4}, \frac{\pi}{2})}\right)\right]+\left[P\left(\frac{1}{2}_{(\frac{\pi}{4}, \frac{\pi}{2})}\right)-P\left(-\frac{1}{2}_{(\frac{\pi}{4}, \frac{\pi}{2})}\right)\right]\right\}\nonumber \\
&+\frac{\sqrt{2}}{24}\left\{27\left[P\left(\frac{3}{2}_{(-\frac{\pi}{4}, \frac{\pi}{2})}\right)-P\left(-\frac{3}{2}_{(-\frac{\pi}{4}, \frac{\pi}{2})}\right)\right]+\left[P\left(\frac{1}{2}_{(-\frac{\pi}{4}, \frac{\pi}{2})}\right)-P\left(-\frac{1}{2}_{(-\frac{\pi}{4}, \frac{\pi}{2})}\right)\right]\right\},\\
S_{L T T}^{x x}=& \frac{\sqrt{2}}{12}\left\{27\left[P\left(\frac{3}{2}_{(\frac{\pi}{4}, 0)}\right)-P\left(-\frac{3}{2}_{(\frac{\pi}{4}, 0)}\right)\right]+\left[P\left(\frac{1}{2}_{(\frac{\pi}{4}, 0)}\right)-P\left(-\frac{1}{2}_{(\frac{\pi}{4}, 0)}\right)\right]\right\} \nonumber\\
&-\frac{\sqrt{2}}{12}\left\{27\left[P\left(\frac{3}{2}_{(-\frac{\pi}{4}, 0)}\right)-P\left(-\frac{3}{2}_{(-\frac{\pi}{4}, 0)}\right)\right]+\left[P\left(\frac{1}{2}_{(-\frac{\pi}{4}, 0)}\right)-P\left(-\frac{1}{2}_{(-\frac{\pi}{4}, 0)}\right)\right]\right\} \nonumber \\
&-\frac{\sqrt{2}}{12}\left\{27\left[P\left(\frac{3}{2}_{(\frac{\pi}{4}, \frac{\pi}{2})}\right)-P\left(-\frac{3}{2}_{(\frac{\pi}{4}, \frac{\pi}{2})}\right)\right]+\left[P\left(\frac{1}{2}_{(\frac{\pi}{4}, \frac{\pi}{2})}\right)-P\left(-\frac{1}{2}_{(\frac{\pi}{4}, \frac{\pi}{2})}\right)\right]\right\} \nonumber\\
&+\frac{\sqrt{2}}{12}\left\{27\left[P\left(\frac{3}{2}_{(-\frac{\pi}{4}, \frac{\pi}{2})}\right)-P\left(-\frac{3}{2}_{(-\frac{\pi}{4}, \frac{\pi}{2})}\right)\right]+\left[P\left(\frac{1}{2}_{(-\frac{\pi}{4}, \frac{\pi}{2})}\right)-P\left(-\frac{1}{2}_{(-\frac{\pi}{4}, \frac{\pi}{2})}\right)\right]\right\} ,\\
S_{L T T}^{x y}=& \frac{1}{12}\left\{2 7 \left[P \left(\frac{3}{2}(0,0)\right)-P\left(-\frac{3}{2}(0,0)\right)\right]+\left[P\left(\frac{1}{2}(0,0)\right)-P\left(-\frac{1}{2}(0,0)\right)\right]\right\}\nonumber\\
&+\frac{1}{12}\left\{2 7 \left[P\left(\frac{3}{2}_{(\frac{\pi}{2}, 0)}\right)-P\left(-\frac{3}{2}_{(\frac{\pi}{2}, 0)}\right)\right]+\left[P\left(\frac{1}{2}_{(\frac{\pi}{2}, 0)}\right)-P\left(-\frac{1}{2}_{(\frac{\pi}{2}, 0)}\right)\right]\right\} \nonumber\\
&+\frac{1}{12}\left\{27\left[P\left(\frac{3}{2}_{(\frac{\pi}{2}, \frac{\pi}{2})}\right)-P\left(-\frac{3}{2}_{(\frac{\pi}{2}, \frac{\pi}{2})}\right)\right]+\left[P\left(\frac{1}{2}_{(\frac{\pi}{2}, \frac{\pi}{2})}\right)-P\left(-\frac{1}{2}_{(\frac{\pi}{2}, \frac{\pi}{2})}\right)\right]\right\} \nonumber\\
&-\frac{\sqrt{2}}{6}\left\{2 7 \left[P\left(\frac{3}{2}_{(\frac{\pi}{4}, 0)}\right)-P\left(-\frac{3}{2}_{(\frac{\pi}{4}, 0)}\right)\right]+\left[P\left(\frac{1}{2}_{(\frac{\pi}{4}, 0)}\right)-P\left(-\frac{1}{2}_{(\frac{\pi}{4}, 0)}\right)\right]\right\} \nonumber\\
&-\frac{\sqrt{2}}{6}\left\{27\left[P\left(\frac{3}{2}_{(\frac{\pi}{4}, \frac{\pi}{2})}\right)-P\left(-\frac{3}{2}_{(\frac{\pi}{4}, \frac{\pi}{2})}\right)\right]+\left[P\left(\frac{1}{2}_{(\frac{\pi}{4}, \frac{\pi}{2})}\right)-P\left(-\frac{1}{2}_{(\frac{\pi}{4}, \frac{\pi}{2})}\right)\right]\right\} \nonumber\\
&-\frac{\sqrt{2}}{6}\left\{27\left[P\left(\frac{3}{2}_{(\frac{\pi}{2}, \frac{\pi}{4})}\right)-P\left(-\frac{3}{2}_{(\frac{\pi}{2}, \frac{\pi}{4})}\right)\right]+\left[P\left(\frac{1}{2}_{(\frac{\pi}{2}, \frac{\pi}{4})}\right)-P\left(-\frac{1}{2}_{(\frac{\pi}{2}, \frac{\pi}{4})}\right)\right]\right\} \nonumber\\
&+\frac{\sqrt{3}}{4}\left\{27\left[P\left(\frac{3}{2}_{(\theta_{x y z}, \frac{\pi}{4})}\right)-P\left(-\frac{3}{2}_{(\theta_{x y z},\frac{\pi}{4})}\right)\right] + \left[P\left(\frac{1}{2}_{(\theta_{x y z},-\frac{\pi}{4})}\right) - \left(-\frac{1}{2}_{(\theta_{x y z},-\frac{\pi}{4})}\right)\right]\right\} ,\\
S_{T T T}^{x x x}=& \frac{1}{4}\left\{2 7 \left[P\left(\frac{3}{2}_{(\frac{\pi}{2}, 0)}\right)-P\left(-\frac{3}{2}_{(\frac{\pi}{2}, 0)}\right)\right]+\left[P\left(\frac{1}{2}_{(\frac{\pi}{2}, 0)}\right)-P\left(-\frac{1}{2}_{(\frac{\pi}{2}, 0)}\right)\right]\right\}\nonumber\\
&-\frac{\sqrt{2}}{8}\left\{27\left[P\left(\frac{3}{2}_{(\frac{\pi}{2}, \frac{\pi}{4})}\right)-P\left(-\frac{3}{2}_{(\frac{\pi}{2}, \frac{\pi}{4})}\right)\right] +\left[P\left(\frac{1}{2}_{(\frac{\pi}{2}, \frac{\pi}{4})}\right)-P\left(-\frac{1}{2}_{(\frac{\pi}{2}, \frac{\pi}{4})}\right)\right]\right\}\nonumber\\
&-\frac{\sqrt{2}}{8}\left\{27\left[P\left(\frac{3}{2}_{(\frac{\pi}{2}, -\frac{\pi}{4})}\right)-P\left(-\frac{3}{2}_{(\frac{\pi}{2}, -\frac{\pi}{4})}\right)\right]+\left[P\left(\frac{1}{2}_{(\frac{\pi}{2}, -\frac{\pi}{4})}\right)-P\left(-\frac{1}{2}_{(\frac{\pi}{2}, -\frac{\pi}{4})}\right)\right]\right\} ,\\
S_{T T T}^{y x x}=&-\frac{1}{4}\left\{27\left[P\left(\frac{3}{2}_{(\frac{\pi}{2}, \frac{\pi}{2})}\right)-P\left(-\frac{3}{2}_{(\frac{\pi}{2}, \frac{\pi}{2})}\right)\right]+\left[P\left(\frac{1}{2}_{(\frac{\pi}{2}, \frac{\pi}{2})}\right)-P\left(-\frac{1}{2}_{(\frac{\pi}{2}, \frac{\pi}{2})}\right)\right]\right\} \nonumber\\
&+\frac{\sqrt{2}}{8}\left\{27\left[P\left(\frac{3}{2}_{(\frac{\pi}{2}, \frac{\pi}{4})}\right)-P\left(-\frac{3}{2}_{(\frac{\pi}{2}, \frac{\pi}{4})}\right)\right]+\left[P\left(\frac{1}{2}_{(\frac{\pi}{2}, \frac{\pi}{4})}\right)-P\left(-\frac{1}{2}_{(\frac{\pi}{2}, \frac{\pi}{4})}\right)\right]\right\} \nonumber\\
&-\frac{\sqrt{2}}{8}\left\{27\left[P\left(\frac{3}{2}_{(\frac{\pi}{2}, -\frac{\pi}{4})}\right)-P\left(-\frac{3}{2}_{(\frac{\pi}{2}, -\frac{\pi}{4})}\right)\right]+\left[P\left(\frac{1}{2}_{(\frac{\pi}{2}, -\frac{\pi}{4})}\right)-P\left(-\frac{1}{2}_{(\frac{\pi}{2}, -\frac{\pi}{4})}\right)\right]\right\} ,
\end{align}
where $\theta_{x y z}$ indicates the angle $\arctan (\sqrt{2})$.
 
The domains of the spin components can be obtained through the probabilistic interpretations
\begin{align}
&S_{L}, S_{T}^x, S_{T}^y \in [-\frac{3}{2}, \frac{3}{2}],\\
&S_{LL} \in [-1,1], 
\quad
S_{LT}^x, S_{LT}^y, S_{TT}^{xx}, S_{TT}^{xy} \in [-\sqrt{3}, \sqrt{3}],\\
& S_{LLL} \in [-\frac{9}{10}, \frac{9}{10}],
\quad
S_{LLT}^x, S_{LLT}^y \in [-\frac{3+\sqrt{21}}{10}, \frac{3+\sqrt{21}}{10}],\\
& S_{LTT}^{xx}, S_{LTT}^{xy} \in [-\sqrt{3}, \sqrt{3}],
\quad
S_{TTT}^{xxx}, S_{TTT}^{yxx} \in [ -3, 3]. 
\end{align}
We note that the ranges may change in accordance with different normalization conventions.

\section{FRAGMENTATION FUNCTIONS AND SYMMETRIC TRACELESS TENSORS} 
\label{a.FFs}

The TMD FFs defined in Eqs.~\eqref{e.tmdff_u}--\eqref{e.tmdff_ttt} can be separated through the projection by the Dirac matrix,
\begin{align}
\Delta^{[\Gamma]}\left(z, k_T\right) &=\operatorname{Tr}\left[\Delta\left(z, k_T\right) \Gamma\right], \\
\Delta^{[\Gamma]}(z) &=\operatorname{Tr}[\Delta(z) \Gamma].
\end{align}
Taking $\Gamma = \gamma^-$, $\gamma^-\gamma_5$, and $i\sigma^{i-}\gamma_5$, one can obtain all leading-twist TMD FFs,
\begin{align}
\Delta_{U}^{\left[\gamma^{-}\right]}\left(z, k_T\right) &=D_{1}, \\
\Delta_{L}^{\left[\gamma^{-}\right]}\left(z, k_T\right) &=0, \\
\Delta_{T}^{\left[\gamma^{-}\right]}\left(z, k_T\right) &=\left(\epsilon_{T}^{\mu \nu} S_{ T \nu} \frac{k_{T \mu}}{M} D_{1 T}^{\perp}\right), \\
\Delta_{L L}^{\left[\gamma^{-}\right]}\left(z, k_T\right) &=S_{ L L} D_{1 L L}, \\
\Delta_{L T}^{\left[\gamma^{-}\right]}\left(z, k_T\right) &=\frac{\boldsymbol{S}_{h L T} \cdot \bm{k}_T}{M} D_{1 L T}^\perp, \\
\Delta_{T T}^{\left[\gamma^{-}\right]}\left(z, k_T\right) &= S_{TT\mu\nu}\frac{k_T^{\mu\nu}}{M^{2}} D_{1 T T}^\perp, \\
\Delta_{L L L}^{\left[\gamma^{-}\right]}\left(z, k_T\right)
&=0,\\
\Delta_{L L T}^{\left[\gamma^{-}\right]}\left(z, k_T\right)
&=\left(\epsilon_{T}^{\mu \nu} S_{ L L T \nu} \frac{k_{T \mu}}{M} D_{1 L L T}^\perp\right),\\
\Delta_{L T T}^{\left[\gamma^{-}\right]}\left(z, k_T\right)
&=\left(\epsilon_{T\nu}^{\mu} S_{ L T T}^{ \nu \rho} \frac{k_{T\mu\rho}}{M^{2}} D_{1 L T T}^\perp\right),\\
\Delta_{T T T}^{\left[\gamma^{-}\right]}\left(z, k_T\right)
&=\left(\epsilon_{T\nu}^{\mu }S_{ T T T}^{ \nu\rho\sigma}\frac{k_{T\mu\rho\sigma }}{M^3}D_{1T T T}^\perp\right),\\
\Delta_{U}^{\left[\gamma^{-} \gamma_{5}\right]}\left(z, k_T\right)&= 0, \\
\Delta_{L}^{\left[\gamma^{-} \gamma_{5}\right]}\left(z, k_T\right)&= S_{ L} G_{1 L}, \\
\Delta_{T}^{\left[\gamma^{-} \gamma_{5}\right]}\left(z, k_T\right)&=\frac{\boldsymbol{S}_{h T} \cdot \boldsymbol{k}_{T}}{M} G_{1 T}^\perp ,\\
\Delta_{L L}^{\left[\gamma^{-} \gamma_{5}\right]}\left(z, k_T\right)&=0 ,\\
\Delta_{L T}^{\left[\gamma^{-} \gamma_{5}\right]}\left(z, k_T\right)&=-\left(\epsilon_{T}^{\mu \nu} S_{ L T \nu} \frac{k_{T \mu}}{M} G_{1 L T}^\perp\right), \\
{\Delta}_{T T}^{\left[\gamma^{-} \gamma_{5}\right]}\left(z, k_T\right)&=\left(\epsilon_{T\nu}^{\mu} S_{ T T}^{\nu \rho} \frac{k_{T\mu\rho} }{M^{2}} G_{1 T T}^\perp\right),\\
\Delta_{L L L}^{\left[\gamma^{-}\gamma_5\right]}\left(z, k_T\right)
&=S_{ L L L}G_{1 L L L},\\
\Delta_{L L T}^{\left[\gamma^{-}\gamma_5\right]}\left(z, k_T\right)
&=\frac{\boldsymbol{S}_{h L L T} \cdot \boldsymbol{k}_{T}}{M} G_{1 L L T}^\perp,\\
\Delta_{L T T}^{\left[\gamma^{-}\gamma_5\right]}\left(z, k_T\right)
&={S}_{h L T T \mu\nu} \frac{{k}_{T}^{\mu\nu}}{M^{2}} G_{1 L T T}^\perp,\\
\Delta_{T T T}^{\left[\gamma^{-}\gamma_5\right]}\left(z, k_T\right)
&=S_{ T T T \mu\nu\rho} \frac{k_{T}^{\mu\nu\rho}}{M^3}G_{1T T T}^\perp,\\
\Delta_{U}^{\left[i \sigma^{i-} \gamma_{5}\right]}\left(z, k_T\right) &=\left(\frac{\epsilon_{T}^{i j} k_{T j}}{M} H_{1}^{\perp}\right), \\
\Delta_{L}^{\left[i \sigma^{i-} \gamma_{5}\right]}\left(z, k_T\right) &=S_{ L} \frac{k_{T}^{i}}{M} H_{1 L}^{\perp} ,\\
\Delta_{T}^{\left[i \sigma^{i-} \gamma_{5}\right]}\left(z, k_T\right) &=S_{ T}^{i} H_{1 T}- \frac{k_{T}^{ij}}{M} S_{Tj} H_{1 T}^{\perp} ,\\
\Delta_{L L}^{\left[i \sigma^{i-} \gamma_{5}\right]}\left(z, k_T\right) &=-\left(S_{ L L} \frac{\epsilon_{T}^{i j} k_{T j}}{M} H_{1 L L}^{\perp}\right) ,\\
\Delta_{L T}^{\left[i \sigma^{i-} \gamma_{5}\right]}\left(z, k_T\right) &=-\left(\epsilon_{T}^{i j} S_{ L T j} H_{1 L T} \right) + \left( \epsilon_{Tj}^{i} \frac{k_{T}^{jl}}{M^2} S_{LTl} H_{1 L T}^{\perp}\right) ,\\
\Delta_{T T}^{\left[i \sigma^{i-} \gamma_{5}\right]}\left(z, k_T\right) &=-\left(\epsilon_{T}^{i j} S_{ T T j l} \frac{k_{T}^{l}}{M} H_{1 T T}^{ \perp}\right)-\left( \epsilon^{i}_{Tj} \frac{k_T^{
jlm}}{M^3} S_{TTlm} H_{1 T T}^{\perp \perp}\right),\\
\Delta_{L L L}^{\left[i \sigma^{i-} \gamma_{5}\right]}\left(z, k_T\right) 
&=S_{ L L L} \frac{ k_{T}^i}{M} H_{1 L L L}^{\perp},\\
\Delta_{L L T}^{\left[i \sigma^{i-} \gamma_{5}\right]}\left(z, k_T\right)
&=S_{ L L T}^i H_{1L L T} -  \frac{ k_{T}^{ij}}{M^2} S_{LLTj}H_{1 L L T}^{\perp},\\
\Delta_{L T T}^{\left[i \sigma^{i-} \gamma_{5}\right]}\left(z, k_T\right) 
&= S_{ L T T}^{i j} \frac{k_{T j}}
{M} H_{1 L T T}^{  ^\perp}+ \frac{k_{T}^{ijl}}{M^3} S_{LTTjl} H_{1 L T T}^{\perp\perp},\\
\Delta_{T T T}^{\left[i \sigma^{i-} \gamma_{5}\right]}\left(z, k_T\right) 
&= S_{T T T}^{i j l} \frac{k_{T jl}}{M^2} H_{1 T T T}^{ ^\perp} + \frac{ k_{T}^{ijlm}}{M^3} S_{TTT jlm}  H_{1 T T T}^{\perp\perp},
\end{align}  
where the terms in parentheses are often referred to as naively T-odd terms and the indices $i$, $j$, $k$, and $l$ indicate the transverse components. The TMD FFs are defined as functions of $z$ and $k_T^2=-\bm{k}_T^2$, with inhomogeneous $k_T$ dependence factored out. The completely symmetric and traceless tensors $k_T^{i_1\cdots i_n}$ are defined as~\cite{Boer:2016xqr}
\begin{align}
k_{T}^{i j}&= k_{T}^{i} k_{T}^{j}-\frac{1}{2} k_{T}^{2} g_{T}^{i j}, \label{e.kij}\\
k_{T}^{i j k}&= k_{T}^{i} k_{T}^{j} k_{T}^{k}-\frac{1}{4} k_{T}^{2}\left(g_{T}^{i j} k_{T}^{k}+g_{T}^{i k} k_{T}^{j}+g_{T}^{j k} k_{T}^{i}\right), \label{e.kijk}\\
k_{T}^{i j k l}&= k_{T}^{i} k_{T}^{j} k_{T}^{k} k_{T}^{l} \nonumber\\
&\quad-\frac{1}{6} k_{T}^{2}\left(g_{T}^{i j} k_{T}^{k l}+g_{T}^{i k} k_{T}^{j l}+g_{T}^{i l} k_{T}^{j k}+g_{T}^{j k} k_{T}^{i l}+g_{T}^{j l} k_{T}^{i k}+g_{T}^{k l} k_{T}^{i j}\right) \nonumber\\
&\quad-\frac{1}{8}\left(k_{T}^{2}\right)^{2}\left(g_{T}^{i j} g_{T}^{k l}+g_{T}^{i k} g_{T}^{j l}+g_{T}^{i l} g_{T}^{j k}\right) , \label{e.kijkl}
\end{align}
which satisfy
\begin{align}
g_{T i j} k_{T}^{i j}=g_{T i j} k_{T}^{i j k}=g_{T i j} k_{T}^{i j k l}=0.
\end{align}


\end{document}